\documentclass{aastex}
\usepackage{emulateapj5,danonecolfloat}

\newcommand{\polar}{{\scshape Polar}}
\newcommand{\polars}{{\scshape Polar's}}

\newcommand{\lan}{\langle}
\newcommand{\ran}{\rangle}
\newcommand{\non}{\nonumber}

\def\ie{{\frenchspacing\it i.e.}}
\def\eg{{\frenchspacing\it e.g.}}

\def\n{{\bf n}}

\def\x{\bf x}
\def\y{\bf y}
\def\z{\bf z}

\def\E{{\bf E}}

\def\bi{\begin{itemize}}
\def\ei{\end{itemize}}

\def\n2{N$_{2}$}
\def\4he{$^{4}$He}
\def\cm3{cm$^3$}

\newcommand{\ka}{{K_{\rm a}}}

\def\beq{\begin{equation}}
\def\eeq{\end{equation}}
\def\beqa{\begin{eqnarray}}
\def\eeqa{\end{eqnarray}}
\def\eq#1{equation~(\ref{#1})}

\received{2001 November 14}
\begin{document}

\twocolumn[

\title{An Instrument for Investigating the Large Angular Scale Polarization of
 the Cosmic Microwave Background}

\author{Brian G. Keating\altaffilmark{1}, Christopher W. O'Dell,
Joshua O. Gundersen\altaffilmark{2}, Lucio
Piccirillo\altaffilmark{3}, Nate C. Stebor\altaffilmark{4}, and
Peter T. Timbie}\affil{Department of Physics, 2531 Sterling Hall,
University of Wisconsin -- Madison, Madison, WI 53706}

\begin{abstract}
We describe the design and performance of a microwave polarimeter
used to make precision measurements of polarized astrophysical
radiation in three microwave frequency bands spanning 26-36 GHz.
The instrument uses cooled HEMT amplifiers in a correlation
polarimeter configuration to achieve high sensitivity and
long-term stability. The instrument demonstrates long term
stability and has produced the most restrictive upper limits to
date on the large angular scale polarization of the 2.7 K cosmic
microwave background radiation.
\end{abstract}


]

\altaffiltext{1}{Current Address: Division of Physics, Math, and
Astronomy, California Institute of Technology, Pasadena, CA 91125;
bgk@astro.caltech.edu.} \altaffiltext{2}{Department of Physics,
University of Miami, Coral Gables, FL 33146}
\altaffiltext{3}{Department of Physics and Astronomy, University
of Wales - Cardiff, Wales, UK CF24 3YB}
\altaffiltext{4}{Department of Physics, University of California
at Santa Barbara, Santa Barbara, CA 93106}

\section{Introduction}

Observations of the cosmic microwave background (CMB) are some of
the most powerful tools in cosmology. The CMB has the promise to
address the most fundamental cosmological questions: the geometry
and age of the universe, the matter content of the universe, the
ionization history and the spectrum of primordial perturbations.
The CMB is specified by three characteristics: its spectrum, the
spatial distribution of its total intensity, and the spatial
distribution of its polarization. All three properties depend on
the fundamental cosmological parameters. Several instruments have
produced precision measurements of its spectrum and anisotropy at
large, medium, and small angular scales (see, for example,
\citet{wan01} and references therein).

Similar to the CMB anisotropy power spectrum, the polarization
power spectrum encodes information on all angular scales.  Large
angular scales ($>1\arcdeg$) correspond to regions on the last
scattering surface which were larger than the causal horizon at
that time. In the absence of reionization, polarization on these
scales was affected only by the longest wavelength modes of the
primordial power spectrum. Reionization is expected to produce a
new polarized peak in the power spectrum near $\ell\la 20$, where
the precise peak location depends on the redshift at which the
Universe \citep{zal98,kea98} became reionized.

The large scale region of the anisotropy power spectrum was
measured by the {\em COBE} DMR, and this established the
normalization for models of large scale structure formation. The
effect of reionization on the anisotropy power spectrum is to damp
all angular scales by a factor of $e^{-2\tau}$ where $\tau$ is the
optical depth to the reionization epoch. This effect is degenerate
with several other cosmological parameters \citep{zal97b} and
non-zero $\tau$ cannot be unambiguously detected, at any scale,
from the anisotropy power spectrum alone. Similarly, the effect of
gravitational waves on the anisotropy power spectrum is also
degenerate with other cosmological parameters \citep{zss97}.
Detection of CMB polarization at scales $>1\arcdeg$ has the
potential to detect reionization and primordial gravitational
waves.

Although the polarization signal at large angular scales is
expected to be weaker than at small scales, the design of a large
angular scale experiment is simpler and more compact than an
experiment probing small scales. A large angular scale experiment
with no external beam forming optics (\ie, no primary mirror),
exhibits minimal spurious polarization and reduces susceptibility
to numerous sources of systematic error. In this paper we describe
our approach to measuring the large scale polarization of the CMB:
Polarization Observations of Large Angular Regions (\polar).

\polars\ design builds on techniques developed in previous
searches for CMB polarization \citep{nan79,lub81,wol97} and is
driven by the size and angular scale of the anticipated CMB
signals, spectral removal of foreground sources, optimization of
the observing scheme, long-term stability, and immunity to
potential systematic effects. \polar\ is a wide bandwidth ($\sim
8$ GHz) correlation polarimeter dedicated to measurements of the
CMB. \polar\ measures polarization in the $\ka$ band, between 26
and 36 GHz, using cooled High Electron Mobility Transistor (HEMT)
amplifiers. This band is multiplexed into three sub-bands to allow
for discrimination against foreground sources. The radiometer
executes a zenith drift scan with a $7\arcdeg$ FWHM beam produced
by a corrugated feed horn antenna. In the Spring of 2000 \polar\
observed a $\simeq 7\arcdeg$ wide region from $RA = 112\arcdeg$ to
$275\arcdeg$ at declination $43\arcdeg$ for 45 days from the
University of Wisconsin -- Madison's Pine Bluff Observatory in
Pine Bluff, Wisconsin (Latitude +43\arcdeg01\arcmin, Longitude
+89\arcdeg45\arcmin). In a single night of data \polar\ achieved a
sensitivity level of $\sim 50 \,\mu$K to the Stokes parameters Q
and U in each beam-sized pixel. For the 2000 season \polar\ set
upper limits on the amplitude of the cosmological E-mode and
B-mode \citep{zal97,kks97} power spectra of $T_E,T_B < 10\,\mu$K
at 95\% confidence \citep{kea01}.

In this paper we describe the design and performance of \polar. In
section \ref{s:corr_pol} we outline the fundamentals of the
correlation polarimeter. Section \ref{s:Rx} presents detailed
instrument design specifications and performance. Section
\ref{s:calib} describes our calibration technique, and section
\ref{s:systematics} addresses potential systematic effects and
radiometric offset characterization. Finally, section \ref{s:cuts}
summarizes the meteorological conditions encountered during the
observation run as well as our data selection criteria.

\section{Correlation Polarimeter}
\label{s:corr_pol}

The correlation polarimeter is based on a correlation radiometer
\citep{fuj64,roh96}, which shares many technological features with
an interferometer \cite{tho98}. The development of the correlation
radiometer preceded the discovery of the CMB in 1965; see for
example \citet{fuj64}. Several early CMB experiments used
correlation radiometers for anisotropy measurements \citep{che79},
spectral measurements \citep{joh87}, and the first application of
a dedicated interferometer to CMB research \citep{tim90}.

\begin{figure}[h]
\hspace{-.0  cm} \centerline{\epsfxsize=9.8cm\epsffile{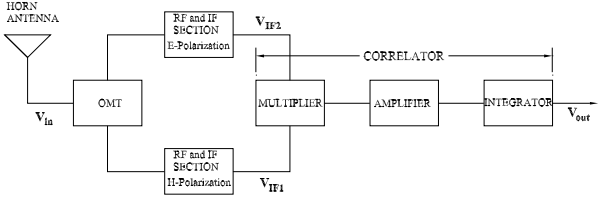}}
\vspace{-0cm}\caption[Correlation Radiometer Schematic] {Schematic
of a simple correlation polarimeter. Radio-frequency fields are
split into two linear polarization states by an orthomode
transducer (OMT), and amplified. The rectangular waveguide output
ports of the OMT define the perpendicular E and H planes of the
polarimeter. The field amplitudes are multiplied, producing a DC
voltage proportional to their product. The DC product voltage is
filtered and amplified before being integrated (low-pass filtered)
prior to being recorded. \label{f:corrrad}}
\end{figure}

In a correlation polarimeter the two linear polarization
components of an incident electric field are separated and
processed in two discrete radiometer chains. For an electric field
propagating in the $\hat{\bf{z}}$ direction, at an electromagnetic
frequency $\nu$, $\E(\nu,t) = E_x(t) \cos (2\pi \nu t + \phi_x)
\hat{\x} + E_y(t) \cos (2\pi \nu t + \phi_y) \hat{\y}$, the Stokes
parameters are defined by: \beqa
I \equiv E_x^2(t) + E_y^2(t) \\
Q \equiv E_x^2(t) - E_y^2(t) \\
U \equiv \lan 2 E_y(t) E_x(t) \cos (\phi_x - \phi_y) \ran \label{e:u}\\
V \equiv \lan 2 E_y(t) E_x (t)\sin (\phi_x - \phi_y) \ran \eeqa
where the angular brackets denote the expectation value or
time-average. For completely linearly polarized radiation, $U =
\lan 2 E_y(t) E_x(t) \ran$ and $V = 0.$ A simplified correlation
polarimeter is shown in figure \ref{f:corrrad}. In a correlation
polarimeter the incident field collected by the feedhorn is
decomposed into two orthogonal polarization spatial modes, defined
by the rectangular waveguide axes of the polarization diplexer
(orthogonal mode transducer -- OMT). The fields, now properly
described as voltages, are amplified in separate amplifier chains
and the output voltages are multiplied and averaged, resulting in
a signal proportional to the $U$ Stokes parameter (equation
\ref{e:u}). As the polarimeter is rotated about the $\hat{\z}$
axis by an angle $\theta$ the correlator output varies as
$R(\theta) \propto U \cos 2\theta + Q \sin 2 \theta$.

Physically, the correlator (multiplier) is based on a diode bridge
which acts as a mixer when provided with an AC bias waveform. The
uncorrelated RF power in each arm provides the bias power for the
diodes. An advantage of this detection mechanism is that probes
the RF fields at frequencies much higher than the $1/f$ knee of
the RF amplifiers. It requires no moving parts which complicate
CMB experiments that spatially modulate the incoming signals to
overcome low frequency drifts. A variation of the correlation
polarimeter design correlates left and right circular polarization
modes, and is able to recover all four Stokes parameters
simultaneously\footnote{A disadvantage of the circular mode
correlation polarimeter is that conversion between $Q$ and $U$ can
occur, whereas for linear correlation polarimeters (such as
\polar) primarily conversion between $U$ and $V$ (or $Q$ and $V$)
occurs \citep{tho98,car01}. Since $V\ll Q,U$ (either cosmological
or systematic), polarization conversion is negligible for \polar.
However, both correlation polarimeter methods suffer from
conversion between $I$ and $Q$ and $U$ as discussed below.}
\citep{sir97,car01}.

The electric field entering the feedhorn at time at time $t$, in
polarization state $i\in\{x,y\}$ from a source in direction
$\theta$ (with respect to the feed boresight axis) is expanded
into: \beqa
\tilde{E}_i(\theta,\nu) = \int_{-\infty}^{+\infty} E(\theta,t) e^{-i 2\pi \nu t - \phi_i} dt \non \\
E_i(\theta,t) = \int_{-\infty}^{+\infty} \tilde{E}(\theta,\nu)
e^{i 2\pi \nu t + \phi_i} d\nu \non .\eeqa The $x,y$ coordinate
basis is defined by the orthogonal $E$ and $H$ output waveguide ports of the OMT.

The antenna output voltage for polarization state $i$ is \beq
\tilde{V}_{i}(\nu)=2\pi \int_{-\pi}^{+\pi} \tilde{E}_i(\theta,\nu)
\tilde{G}(\theta,\nu) d\theta,\eeq where $\tilde{G}(\theta,\nu)$
is the horn's (axisymmetric) voltage response function. The output
voltage for each polarization, after amplification with total radiometer voltage
transfer function $\tilde{H}(\nu)$, is ${V\prime}_i(\nu) =
\tilde{H}(\nu)\tilde{V}_{i}(\nu)$. The correlator produces the
\emph{complex} correlation function, at time lag $\tau$, of the
two voltages: \beqa R(\tau) =\lim_{T \rightarrow \infty}
\frac{4\pi^2}{2T}
\int_{-T}^{+T} V_x(t)V^*_y(t-\tau) dt\non\\
=\lim_{T \rightarrow \infty} \frac{4\pi^2}{2T} \int_{-T}^{+T} dt
\int_{-\infty}^{+\infty} d\nu_x \int_{-\infty}^{+\infty} d\nu_y
\int_{0}^{\pi} d\theta_x \int_{0}^{\pi} d\theta_y  \non \\
\times\, \tilde{E}_x(\theta,\nu_x)\tilde{E}^*_y(\theta',\nu_y)
\tilde{H}_x(\nu_x)\tilde{H}^*_y(\nu_y)
\tilde{G}_x(\theta_x,\nu_x)\tilde{G}^*_y(\theta_y,\nu_y)\non\\
\times\, e^{i(2\pi \nu_x t +\phi_x)}e^{-i(2\pi \nu_y (t-\tau)
+\phi_y)}\hspace{10pt}. \eeqa

Remembering that $\int_{-\infty}^{+\infty} e^{2\pi i\, t(\nu_x -
\nu_y )} dt = \delta(\nu_x - \nu_y)$, we obtain: \beqa R(\tau) =
4\pi^2\int_{-T}^{+T}\int_{0}^\pi \int_{0}^\pi d\nu
\, d\theta_x \, d\theta_y \hspace{3cm}\non \\
\times\, \tilde{\gamma}(\nu,\theta_x,\theta_y)
\tilde{B}(\nu,\theta) \tilde{H}_x(\nu)\tilde{H}^*_y(\nu)e^{i(2\pi
\nu  \tau +\Delta \phi)}, \eeqa where: \beq
\tilde{\gamma}(\nu,\theta_x,\theta_y)  = \lim_{T \rightarrow
\infty}
\frac{1}{2T}[\tilde{E}_x(\theta,\nu)\tilde{E}^*_y(\theta',\nu)]
\eeq is the \emph{source coherence function}, $\Delta \phi =
\phi_x-\phi_y$, and
$$\tilde{B}(\nu,\theta_x,\theta_y) = \tilde{G}_x(\theta_x,\nu)
\tilde{G}^*_y(\theta_y,\nu).$$  In practice, it is not necessary
to enforce $T \rightarrow \infty$, as long as $T \gg 1/\nu$. If,
as is the case for \polar,
$\tilde{G}_x(\theta_x,\nu)\simeq\tilde{G}_y(\theta_y,\nu)\equiv
\tilde{G}(\theta,\nu)$, then ${B}(\nu,\theta)=\vert
\tilde{G}(\theta,\nu)\vert^2$ is the power response function of
the horn, or \emph{beam pattern.}  For a thermal source, such as
the CMB,
$\tilde{\gamma}(\nu,\theta,\theta')=\tilde{\gamma}(\nu,\theta)\delta(\theta-\theta')$.
For \polar\ $\tilde{H}_x(\nu)\simeq\tilde{H}_y(\nu)\equiv
\tilde{H}(\nu)$, and only the real part of the complex correlation
function is measured with zero lag. Thus, \polars\ output can be
expressed as: \beq\label{e:coherence} R_o =
4\pi^2\int_{-\infty}^{+\infty}d\nu \int_{0}^{+\pi}d\theta \;
\tilde{\gamma}(\nu,\theta) \tilde{B}(\nu,\theta) \vert
\tilde{H}(\nu)\vert^2 \, \cos{(\Delta\phi_\nu)} \eeq where the
$\nu$ subscript on $\Delta\phi$ incorporates a (potentially)
frequency dependent phase shift between the two arms of the
radiometer; see section \ref{ss:phaseshift}. The properties of the
source coherence function, the transfer functions, and the beam
response completely determine the output voltage. Equation
\ref{e:coherence} will be useful in section \ref{s:calib} where
\polars\ response to completely correlated, polarized signals
produced by calibration sources is computed.

\subsection{Minimum Detectable Signal}

The sensitivity of the correlation polarimeter depends on both the
system noise temperature and the RF bandwidth of the system. Since
there are two RF amplifier chains, the system temperature is their
geometric mean: $T_{sys} = \sqrt{T_{sys}^E T_{sys}^H}$, and the
minimum detectable signal in an integration time $\tau$ is: \beq
\label{eq:truedTrms} \Delta T = \sqrt{\frac{2 T_{sys}^E
T_{sys}^H}{\Delta\nu_{RF}\,\tau\,\cos^2{\lan\Delta\phi\ran_\nu}}},
\eeq where $\Delta\nu_{RF}$ is the RF bandwidth, and $E$ and $H$
refer to the orthogonal polarization states separated by the OMT
\citep{tho98}. Here, we define $\lan\Delta\phi\ran_\nu$ as the
band-averaged differential phase shift between the two arms prior
to correlation: $\int_{0}^\infty \Delta \phi(\nu)
d\nu/\int_{0}^\infty \Delta \phi^2(\nu) d\nu$. Depending on the
physical nature of the differential phase shift (e.g.,
differential guide length, dielectric or microstrip dispersion,
etc.) the phase shift may be calculable and/or separable from
other sources of bandwidth degradation. In POLAR, all passive
waveguide components are matched in length, and adjustable coaxial
phase shifters prior to correlation are employed to minimize the
contribution of path length phase differences. Equation
\ref{eq:truedTrms} is valid only when there are no RF gain or
offset fluctuations. The effects of radiometer offset, stability,
and other non-idealities are discussed in section
\ref{s:systematics}.

\section{The \polar\ Radiometer}
\label{s:Rx} \polars\ radiometer is comprised of three sections:
1) cold receiver components: optics, OMT, isolators, HEMT
amplifiers, 2) room-temperature receiver components: warm RF
amplifiers, heterodyne stage, warm IF amplifiers, band-defining
filters, correlators, and 3) post-detection components:
pre-amplifiers, low frequency processing, and data acquisition. In
this section the details of the experimental design are presented.

\begin{figure}[t]
\hspace{-.0cm} \centerline{\epsfxsize=9.8cm\epsffile{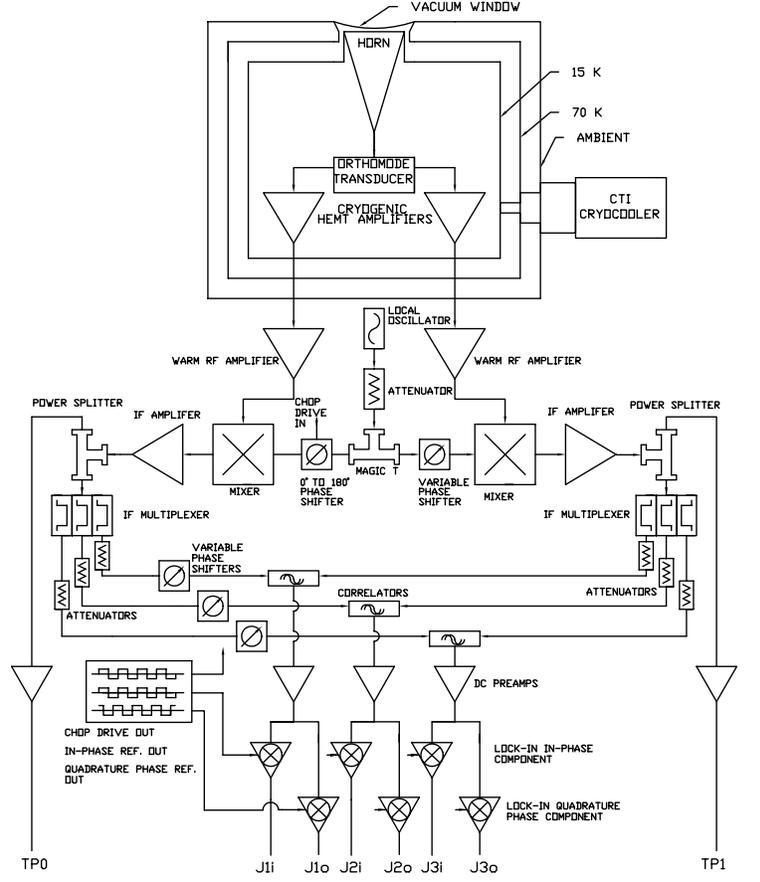}}
\vspace{-0cm}\caption[Components of the \polar\ $\ka$ Band
Radiometer] {Schematic of the \polar\ $\ka$-band correlation
polarimeter. \label{f:signalchain}}
\end{figure}

\begin{deluxetable}{ccccccc}
\tablewidth{0pt}\hspace{-4cm}\tablecaption{\polar\ Instrument
Specifications\label{t:polarspex}} \tablehead{ &
\colhead{$\nu_c$\tablenotemark{b}}&
\colhead{$\Delta\nu$\tablenotemark{c}}&
\colhead{\small{$\Omega_B$} \tablenotemark{d}}&
\colhead{$\overline{T_{\rm{pol}}}$\tablenotemark{e}}&
\colhead{$S_{\rm{sky}}$\tablenotemark{f}}&
\colhead{$T_{\rm{rec}}$\tablenotemark{g}}\\
\colhead{\small{Channel}\tablenotemark{a}} &
\colhead{[\footnotesize{GHz}]} & \colhead{[\footnotesize{GHz}]} &
\colhead{[$\arcdeg$]}&\colhead{[\small{$\mu$K}]}&\colhead{[\footnotesize{$\rm{mK}\,s^{1/2}$}]}&\colhead{\footnotesize{[K]}}\\
} \startdata
\small{TP-E/H} & 31.9 & \small{7.8/8.0}&  7.0 &  \nodata & \small{14.0/20.0}& 32  \\
J3 & 27.5 &  1.3 & 7.5 &  84(28) & 2.0 & 43\\
J2 & 30.5 & 3.1 &  7.0 &  72(14) & 1.2 & 34\\
J1 & 34.0 &3.1 & 6.4 &  33(11) & 1.1  & 33 \\
\enddata
\tablenotetext{a}{\footnotesize{TP-E and TP-H measure the total
power in the E and H polarization planes of the Horn/OMT assembly
prior to correlation.}} \tablenotetext{b}{\footnotesize{Channel
band centroid measured with swept, coherent source.}}
\tablenotetext{c}{\footnotesize{Measured channel bandwidth.}}
\tablenotetext{d}{\footnotesize{Beamwidths (FWHM). E and H-plane
Beamwidths are equal to within $1\%$. Measured feed/OMT
cross-polarization is $<-40$dB for all channels.}}
\tablenotetext{e}{\footnotesize{Mean polarized offset for the
Spring 2000 observing season.
~{$\overline{T_{\textrm{pol}}}=\sqrt{\overline{Q}^2+\overline{U}^2}$}\,
where $\overline{Q}$ and $\overline{U}$ are the average Stokes
parameter offsets for the season. Numbers in parentheses denote
the corresponding values for the QPC.}}
\tablenotetext{f}{\footnotesize{Measured channel NET for a typical
clear day with $K_{\rm{a}}$-band zenith sky temperature $T_{Atm}
\simeq 12$ K. NET measured at Stokes modulation frequency $0.065$
Hz.}}\tablenotetext{f}{\footnotesize{Measured receiver temperature
for each channel.}}
\end{deluxetable}

\subsection{Cryogenics}

\begin{figure}[t]\hspace{-0cm}
\centerline{\epsfxsize=9.8cm\epsffile{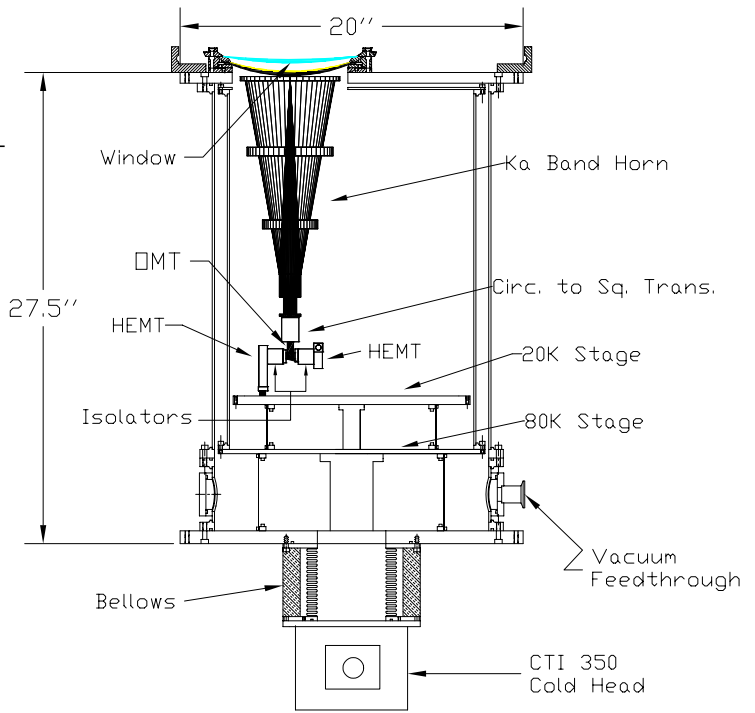}}
\hspace{-5cm}\vspace{1cm}\caption[\polar Dewar]{\polar\ dewar and
$\ka$-band cold receiver components. The horn is located off of
the symmetry axis of the dewar in order to allow for future
receivers to perform simultaneous observations. The HEMT outputs
travel via stainless steel waveguides to a flange on the dewar
bottom.\label{f:dewar}}
\end{figure}

The dewar (figure \ref{f:dewar}) was custom
fabricated\footnote{Precision Cryogenic Systems: Indianapolis, IN}
to house a cryocooler coldhead, and is large enough to accommodate
possible upgrades including additional feed horns in the nominal
20 K (second stage) working volume. The first stage is used to
cool a radiation shield, which is maintained at a nominal
temperature of $\sim 80$ K.

Following pump-down to $\sim 1\times 10^{-4}$ Torr, the pump is
detached and the cryocooler's compressor (CTI 8500 Air Cooled) is
activated. In the field it was found that the ultimate cold stage
temperatures are correlated with the ambient temperature of the
shelter in which \polar\ resides. The compressor is air-cooled;
water cooling was not possible due to the receiver's continuous
rotation. The air cooling causes the compressor's compression
ratio to be a function of ambient temperature, which
modifies its cooling efficiency. Maintaining the temperature
stability of the compressor is accomplished, to first order, by a
commercial air-conditioner during the summer months which counters
the $\sim 2$ kW heat output from the compressor. During the
winter, the heat output by the compressor kept the enclosed
\polar\ shelter at a nearly constant temperature. The compressor
is mechanically isolated from the radiometer by use of a separate
rotation bearing coupled loosely to the motor-driven main bearing
by copper braid (see figure \ref{f:polarcube}). The compressor is
further isolated on its bearing by use of rubber padding on all
support structures.

The cold radiometer components are mounted on the 20 K stage,
located inside the 80 K stage radiation shield. Both waveguide
outputs from the HEMTs connect to vacuum-tight WR-28 stainless
steel waveguide feedthroughs\footnote{Aerowave Corp., Medford, MA}
on the 300 K dewar wall. The feedthroughs are mounted on a single
flange, which also serves as a feedthrough for the HEMT bias
wiring and the temperature diode readout wiring. The final major
port in the dewar is the main vacuum optical window. This port is
located $\sim 3$ inches radially off the rotation axis of the
cryostat to allow for additional feed horns at higher frequency.

\subsection{Optics}
\polars\ RF optical system is composed of a single corrugated feed
horn. Due to the absence of supplemental beam-forming reflectors,
cross-polarization of the instrument is near the minimum possible
level for a millimeter wave receiver. \polars\ feed horn design is
based on the procedure outlined in \citet{zha93}, and is similar
to the $\ka$-band feed horn employed by the \emph{COBE} DMR
experiment \citep{jan79}. The horn exhibits symmetry between its E
and H planes and produces a diffraction-limited power response
with a $\simeq 7\arcdeg$ full-width-at-half-maximum (FWHM) across
the band.

\begin{figure}[t] \hspace{-.5cm}
\centerline{\epsfxsize=10.8cm\epsffile{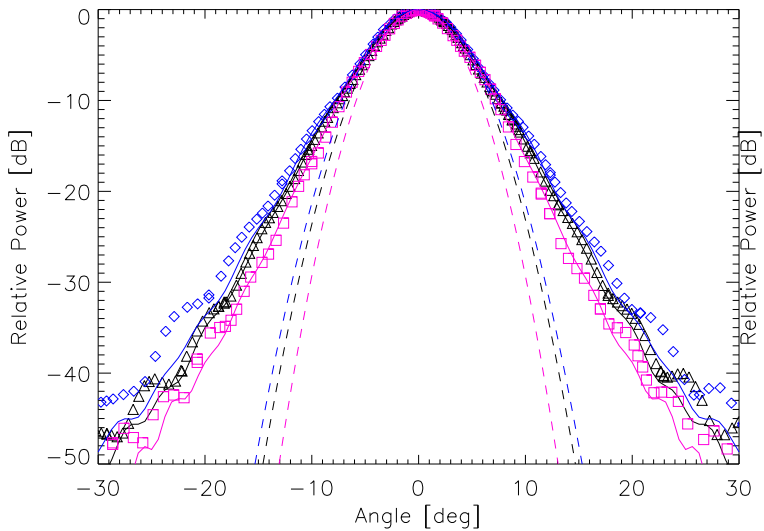}}\caption
{Gauss-Laguerre and Gaussian beam models compared to measured beam
patterns. The diamonds (26 GHz), triangles (29 GHz), and squares
(36 GHz) are the measured beam patterns in the E-plane. The solid
lines represent the corresponding Gauss-Laguerre approximations,
and the dashed lines are the best-fit Gaussians to the main beam
of the measured data. \label{f:e29gltheoryandexpt}}
\end{figure}

\begin{figure}[t] \hspace{-.5cm}
\centerline{\epsfxsize=10.8cm\epsffile{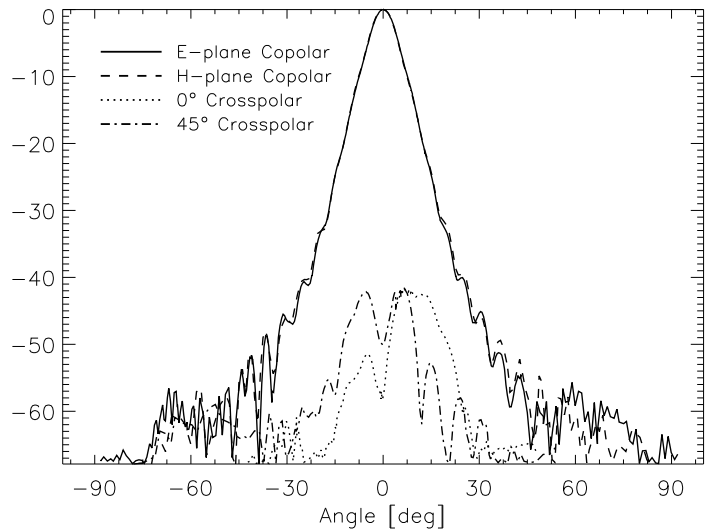}}\caption[29 GHz
Cross-Polarization Beam Map] {Beam maps of the feed-horn measured
at 29 GHz are shown. The solid line is the co-polar E-plane power
response pattern, the dashed line is the co-polar H-plane pattern,
the dotted line is the E-plane cross-polarization response, and
the dot-dashed line is the cross-polarization measured at
$45\arcdeg$ to the E-plane. \label{f:crosspolbeammap}}
\end{figure}

The beam pattern for the \polar\ feed was computed using an 11
term Gauss-Laguerre model ~\citep{cla84} to predict the far-field
beam pattern out to $\sim 20\arcdeg$. A comparison of the measured
and modelled beams is illustrated in figure
\ref{f:e29gltheoryandexpt}. The simple Gauss-Laguerre model breaks
down at low-power levels, which translates to the far off-axis
response of the horn at $\theta \geq 20\arcdeg$. In the absence of
a reliable model for the far off-axis behavior of our feed, we
measured the beam response for a variety of frequencies, for
both polarizations, as well as the cross-polarization response
(see figure \ref{f:crosspolbeammap}).

The final component of the feed-horn is the mode converter, which
is a separate electroformed element at the throat of the horn. The
mode converter combines the TE$^\circ_{11}$ and TM$^\circ _{11}$
circular waveguide modes to create the hybrid HE$^\circ _{11}$
corrugated waveguide mode ~\citep{cla84, zha93}. Following the
throat in the optical path, there is an electroformed transition
from the throat's circular output waveguide to the square-input
waveguide of the OMT. This device was designed by matching the
cutoff wavelengths of the TE$^{\Box}_{10}$ and the HE$^\circ_{11}$
modes.

\subsection{Orthomode Transducer (OMT)}
The OMT (variously referred to as a: polarization diplexer,
dual-mode transducer, orthomode tee, and orthomode junction) is a
waveguide device used to separate the two orthogonal linear
polarization states. \polars\ OMT \footnote{Atlantic Microwave:
Bolton, MA, Model OM 2800} is a three-port device with a square
input port, and two rectangular output ports containing the
orthogonal polarization signals.

The OMT's entrance port is $\ka$-band square guide which
simultaneously supports both $TE^{\Box}_{01}$ and $TE^{\Box}_{10}$
modes. After the modes are separated by the OMT they are further
isolated using cryogenic $\ka$-band isolators\footnote{Pamtech
Corporation: Camarillo, CA}. The isolators prevent coupling of the
polarization states from reflection by high-VSWR components (such
as the HEMT amplifiers). After leaving the OMT the fields in each
of the two polarization states are amplified, downconverted, and
filtered separately until correlation. \polars\ OMT can be
described by a $4\times4$ scattering matrix, $\bf{S}_{\rm{ij}}$.
Element 1 in the S-matrix refers to the input port with E-plane
polarization while 2, 3, 4 refer to input H-plane, output E-plane,
and output H-plane respectively. On-diagonal elements of
$\bf{S}_{\rm{ij}}$ such as $\bf{S}_{\rm{11}}$ and
$\bf{S}_{\rm{22}}$ define the return loss for input E-plane and
H-plane polarization states. The terms $\bf{S}_{\rm{13}}$ and
$\bf{S}_{\rm{24}}$ determine the co-polar transmission/forward
loss, and are not necessarily equal. Differential loss (\eg,
$\bf{S}_{\rm{13}}\neq \bf{S}_{\rm{24}}$ will lead to instrumental
polarization and/or depolarization. The off-diagonal terms
$\bf{S}_{\rm{34}}=\bf{S}_{\rm{43}}$ characterize the output
polarization isolation, and the terms $\bf{S}_{\rm{14}}$ and
$\bf{S}_{\rm{23}}$ define the OMT's cross-polarization. Plots of
the OMT performance are displayed in figure \ref{f:omtquad}. As
described in section \ref{s:systematics}, an offset in the output
of the correlation polarimeter can be caused by either non-zero
cross polarization or isolation.

\subsection{Signal Processing}
\polars\ High Electron Mobility Transistor (HEMT) amplifiers
\citep{pos92}  provide a gain of $\sim 30$ dB. \polars\ amplifiers
utilize InP based devices for the first stage (which have lower
noise-temperatures than GaAs devices) at the expense of slightly
increased $1/f$ noise. However, the low-frequency spectral
properties of these amplifiers are largely irrelevant for
correlation radiometers since the multiplication is performed at
several-GHz, i.e. well above the few-Hz knee of the HEMTs.
\polars\ two amplifiers have noise temperatures of $\simeq 30$ K.
Figure \ref{f:allpsd} shows the low frequency power spectra of the
total power radiometer channels (dominated by the HEMTs) compared
to the spectra of the correlator channels.

\subsubsection{Room Temperature Radiometer Box (RTRB)}
After amplification by HEMTs, the RF signals
are routed to straight 6 inch long stainless steel
waveguides which provide a thermal break from the 20 K HEMTs to
the 300 K dewar walls. The stainless guides are bolted to the
vacuum-tight WR-28 waveguide feedthroughs. Outside the dewar,
straight sections of rhodium plated, brazed-copper waveguides are
used to compensate for the path-length differences between the two
polarizations incurred by the bends. Finally, the waveguides
enter the room temperature radiometer box (RTRB), where the
signals are converted from waveguide to coax to match the inputs
of the $\ka$-band warm HEMT amplifiers\footnote{MITEQ: Hauppauge,
NY, Model JS426004000-30-8P} . The noise temperatures of these
devices are $T_N \simeq 230$ K.

\subsubsection{Superheterodyne Components}
Following the second-stage of amplification, the signals are
down-converted in frequency from 26-36 GHz to 2-12 GHz by a 38 GHz
local oscillator\footnote{Millimeter Wave Oscillator Co.:
Longmont, CO} (LO) and superheterodyne mixers\footnote{MITEQ:
Model TB0440LW1}. The IF spectrum is a (scaled) replica of the
input RF band, with a nearly identical bandwidth. Two stages of IF
(2-12 GHz) amplification are used\footnote{MITEQ: Model
AFS6-00101200-40-10P-6} to provide the appropriate bias power
level into the multipliers. The gain of the IF amplifiers fall
steeply above a frequency of $f_{3dB} \simeq 12$ GHz. Since each
multiplier requires $\sim 5$ mW of bias power, the IF signal must
be amplified by $\sim 60$ dB. After mixing and IF amplification
the signals are divided into two paths. One path, referred to as
the `total power detector' channels (TP-E and TP-H) is detected by
Schottky diodes\footnote{Hewlett Packard: Model HP 8474C}. `E' and
'H' refer to the OMT port in which the respective total power
signal originates. The other post-IF gain path is sent into a
frequency triplexer\footnote{Reactel Corp.: Gaithersburg, MD}. The
function of the triplexers is three-fold. First they produce three
(ideally) independent bands with which are used to investigate the
spectral behavior of the data. Secondly, these devices allow us to
flatten the gain of the system across the wide RF-bandwidth
provided by the HEMTs. Finally, the differential phase between the
two arms can be made flatter across the sub-bands than across the
full RF band. Prior to correlation the gain and phase of each
sub-band are matched with fixed attenuators and phase
shifters\footnote{Weinschel Corp.: Model 917-22}.

\subsubsection{Correlators}
\polars\ correlator is a Schottky-diode mixer\footnote{MITEQ:
Model DBP112HA}. A mixer-based correlator is composed of a double
balanced mixer, a phase modulating element, and lock-in detection.
The primary difference between a multiplier and a conventional
mixer is that the IF bandwidth of the multiplier is made
intentionally narrow to suppress frequency components greater than
$\sim 100$ MHz, and the output of the multiplier can support DC.

The RF band passes of the multipliers are from 1-12 GHz and the IF
bandpasses are from 0-100 MHz. The IF output port is not
transformer-coupled, and propagates the DC signal
proportional to the correlation between signals in the $x$ and $y$
polarization states. Phase modulation and phase-sensitive
detection (PSD) is accomplished by an electronic
$0\arcdeg-180\arcdeg$ phase shifter\footnote{Pacific Millimeter
Products: Golden, CO}, and a synchronous
demodulator\footnote{Analog Devices: Model AD 630} and integrator.
The phase of the LO is switched between $0\arcdeg$ and
$180\arcdeg$ at 1 kHz prior to mixing the $E^{RF}_y$ waveform. The
voltage produced by the correlators at this stage switches between
$\kappa E^{RF}_x E^{RF}_y $ and $-\kappa E^{RF}_x E^{RF}_y $ at 1
kHz, where $\kappa$ is the intensity-to-voltage conversion factor
determined during calibration (section \ref{s:calib}). The output
of the lock-in detectors is proportional to the correlated
component in each arm of the polarimeter.

Two lock-in detectors per correlator are used: one in-phase with
the phase shifter modulation, and the other for the component
$90\arcdeg)$ out of phase. The latter are referred to as
quadrature phase channels (QPC), and are used as noise monitors as
discussed below. Signals leave the pre-amp card and enter a
separate RF-tight box containing six separate lock-in circuits,
corresponding to phase sensitive detection of three correlators,
each with two reference phases, ``in-phase'' and
``quadrature-phase''. The demodulated signal is low-pass filtered
at 5 Hz.

\subsubsection{Post-Detection Electronics}

The pre-amplifier is the final component of the signal chain for
the total power detectors, and the penultimate component for the
correlators (as these are post-detected via the lock-in circuits
described above). To minimize the susceptibility to
electromagnetic interference (EMI), the signals are amplified and
filtered before leaving the radiometer box. A single circuit board
contains five (two total power channels, three correlator
channels) circuits. The card is mounted $\simeq 3$ inches from the
correlators and shares the same thermally regulated environment.
The first stage of the post-detection electronics is a low noise pre-amplifier.
Following the gain stage is a 4-pole, 5 Hz anti-aliasing filter\footnote{Frequency Devices:
Haverhill, MA}. The bandpass of the anti-aliasing filter also sets the fundamental
integration time, $\tau$.

\subsubsection{Electronics Box and Housekeeping}
Thermal regulation of the RTRB is essential to the stability of
the instrument over long periods of time.  The most temperature
sensitive components are the non-linear devices such as the
mixers, multipliers, and especially the Gunn oscillator. The
temperature coefficient of the oscillator output was
$\sim 1 \,\rm{mW \,K}^{-1}$ and the gain following the oscillator was $\sim 20$ dB.
The correlators required 5 mW of bias power so the oscillator's temperature was kept stable to
$\sim 10\, \rm{mK \, hour^{-1}}$ resulting in a maximum bias power change of
$\sim 0.3\%$ per rotation of the polarimeter. To regulate the temperature, a closed-loop
thermal control circuit using feedback from a sensor inside the RTRB was constructed. This circuit used
a commercial microprocessor-based PID controller\footnote{Omega
Inc.: Stamford, CT}, and was capable of regulating up to 300 W of power applied
directly to six 25 W heater pads\footnote{MINCO: Minneapolis, MN}.
Several other housekeeping signals, including temperature sensor
diodes inside the cryostat (on the HEMTs, 20 K cold plate, and
feed horn) and the dewar pressure are monitored. A multi-stage
power regulation approach is implemented. This system employs
precision voltage regulators and references throughout the RTRB;
all signal circuitry (HEMT bias cards, post-detection electronics,
etc.) are voltage regulated and EMI shielded.

\subsubsection{Data Acquisition}
The data acquisition system is composed of an analog-to-digital
converter\footnote{National Instruments DIO-MIO-16 Daqpad}, and a
notebook computer running National Instruments Labview
software. The 16 bit analog-to-digital converter (ADC) samples all
8 data channels as well as 8 housekeeping channels at 20 Hz. By
digitizing all of the data in close physical proximity ($\simeq
10$ in) to the detectors, potential EMI contamination is reduced.
The rotation angle is indexed by a 12 bit relative angular encoder
and a one-bit absolute angular encoder (once per $360 \arcdeg$
rotation). The data files are indexed by calendar time and date,
with several hundred files stored per day. After 7.5 minutes of
acquisition, the files are transferred from the notebook
computer (located on the rotation platform) to a desktop computer
via a local area network Ethernet connection. The coax Ethernet
connection leaves the rotating electronics box through 2 channels
of a 10 channel shielded slip-ring.

\begin{figure}[t]
\centerline{\epsfxsize=9.8cm\epsffile{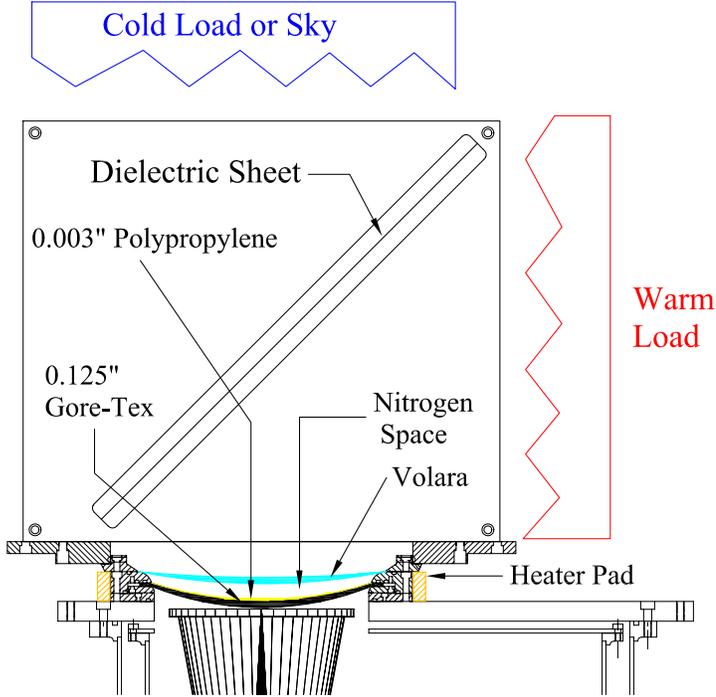}}\caption{Side view
of window and calibrator configuration. The calibrator grids
(either the wire grid or the dielectric sheet - Section
\ref{s:calib}) mount on a rotation bearing attached to the window
flange. The calibrators fill the beam and rotate around the horn
axis. For both the dielectric sheet and the wire grid calibrators,
hot and cold loads are used to generate $\sim 100\%$ polarized
sources, ranging from 20 mK (dielectric sheet) to 200 K (wire
grid). \label{f:window}}
\end{figure}

\subsection{Vacuum Window}
A multi-element vacuum window (figure \ref{f:window}) is composed
of a 0.003 inch vacuum-tight polypropylene vacuum barrier and a
0.125 inch  (permeable) Gore-Tex\footnote{W.L. Gore \& Associates:
Newark, DE} layer which supports the atmospheric load on the
window. A layer of Volara\footnote{Voltek Corp.: Lawrence, MA}
(expanded polyethylene) seals in a dry-nitrogen gas layer between
the polypropylene and prevents condensation and ice on the vacuum
window. A resistive heater element wrapped around the vacuum
window flange warms the window to to $\sim 27$ C to reduce the
formation of dew. With this window the dewar pressure remains
below $10^{-6}$ Torr for months at a time. The emission from the
window is estimated to be $\lesssim 20$ mK, and the reflected power
coefficient is $\lesssim 1\%$.

\subsection{Ground Screens}
\begin{figure}[h]
\hspace{-.5cm}
\centerline{\epsfxsize=9.8cm\epsffile{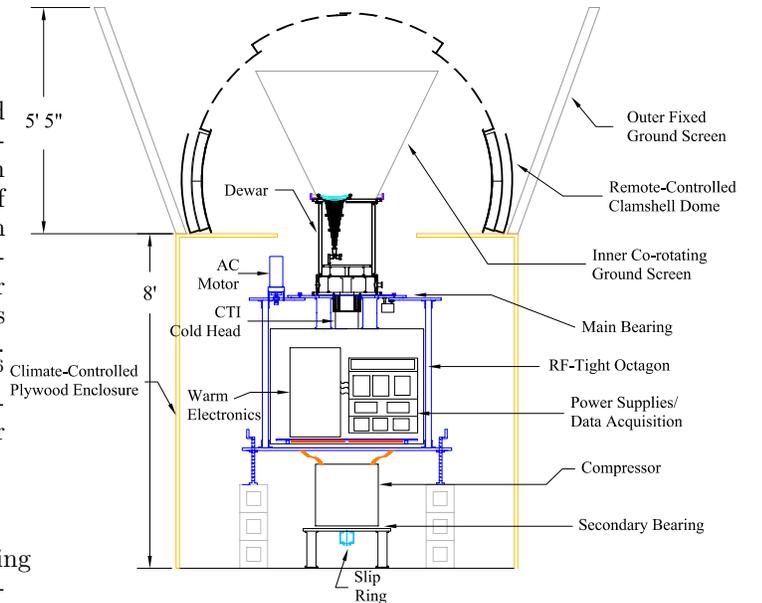}}\caption{\polar\
observatory and ground screens. Two sets of ground screens are
used to reduce the polarized spillover from the earth, as well as
polarized emission from the shields themselves. The outer shield
is fixed to the structure in which \polar\ resides, and is
composed of a lightweight steel skeleton covered by 0.05" aluminum
sheets. The inner ground screen is covered with flat Eccosorb
panels, and co-rotates with the radiometer. Also shown is the
motor-driven, fiberglass clamshell-dome which is remotely operated
via the World Wide Web in the event of inclement weather. The
rotation mount, drive motor, bearing, and angular encoder are also
shown. \label{f:polarcube}}
\end{figure}

\polar\ uses two concentric ground screens; one co-rotating with
the receiver, the other fixed to the observatory structure (see
figure \ref{f:polarcube}). The use of two ground screens is not
unusual in the field, although \polars\ screens are designed to
reject polarized beam spillover, rather than unpolarized,
total-power spillover. The inner ground screens are designed to
terminate the side-lobe power in a known temperature source and
absorb, rather than reflect, solar and lunar light. The inner
conical ground screen is covered with $0.5$ inch Eccosorb foam
designed to suppress specular reflection\footnote{Emerson \&
Cuming: Randolph, MA, Product LS-26}. This absorptive approach is
uncommon in CMB anisotropy experiments as it increases the total
power loading on the detectors. However, the increase in system
temperature due to the inner shield is estimated to be $\lesssim
1$K. The absorption of the foam is greater than 30 dB, and the
estimated induced polarization is estimated to be $<0.5\%$ leading
to a maximum polarization produced by the foam of $< 1$ mK. The
analogous figure for a metallic screen would be ten to one hundred
times larger. Since the inner ground screen co-rotates with the
receiver, it will only produce a constant offset to which the
instrument is insensitive.

The second level of shielding is of the more conventional
reflective-scoop design, $\eg$ \citet{wol97}, and is designed as a
sun-shade for the inner shield. The scoop is mounted to the side
of the \polar\ observatory, and is made of four aluminum panels,
8' wide and 4.8' high.  The level of sidelobe suppression is
estimated using Sommerfeld's scalar diffraction theory for points
deep in the shadow region of a knife-edge scatterer \citep{jac75}.
The estimated suppression is $\simeq -40$ dB, which in combination
with a similar (measured) figure from the inner ground screen, and
the sidelobe response of our feed horn, gives a total estimated
sidelobe suppression better than -100 dB. The response at
$90\arcdeg$ off-axis relative to the peak forward gain was
measured to be $<-50$ dB using a polarized source transmitting in
the $\ka$-band from various locations around the instrument
enclosure. As discussed in section \ref{s:offsets}, the square
shape of the scoop is thought to have produced a $\sim 100\, \mu$K
offset in the Stokes parameters.

\vspace{1cm}

\subsection{Rotation Mount and Drive System}

Measurement of the Stokes parameters is dependent on signal
modulation under rotation. POLAR employs a 30 inch diameter
bearing and AC motor system to rotate the cryostat at 2 RPM ($\sim
33$ mHz). An AC motor produced smoother motion than several
stepper motors tried initially, and was chosen for continuous
rotation. The dewar rides on a bearing composed of two plates each
with a 0.100 inch wide channel filled with $\sim 400$
stainless-steel ball-bearings. The motor pulley has a 12 bit
relative angle encoder which reads out the rotation angle. In
addition, a one-bit absolute encoder is triggered once per
revolution and this defines the zero
angle of the instrument frame. In order to decouple the vibrations
produced by the cryocooler compressor from the receiver, a
separate, vibration isolated rotation mount is used to support the
compressor. The second bearing is loosely coupled to the main
rotation bearing/AC motor drive using braided copper straps. Power
and ethernet connections interface with the rotating system via
the 10 channel slip ring. The mount is not steerable, so \polar\
is restricted to zenith scans.

\subsection{Instrument Bandpasses}
\begin{figure}[b]
\hspace{-.0cm}
\centerline{\epsfxsize=10.cm\epsffile{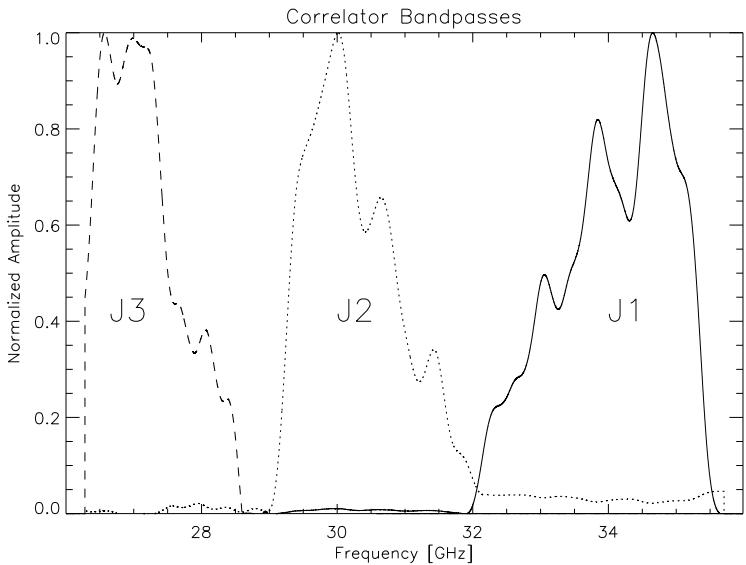}}\caption{Bandpasses
of all three correlator channels are shown, not including the HEMT
amplifier response. A $\ka$-band correlated signal is injected
into each arm of the warm electronics, and the signal frequency is
swept from 26 to 36 GHz to measure the bandpasses.
\label{f:bandpasses}}
\end{figure}

Laboratory measurement of the room temperature radiometer box
bandpasses used an HP 83751A Synthesized Sweeper and an active frequency doubler
to produce a swept signal from from 26 - 36 GHz and fed
into a power splitter. The outputs from the power splitter were
100\% correlated, and these signals were fed into the waveguide
input ports of the RTRB.  The bandpasses of the three correlator channels are shown in
figure \ref{f:bandpasses} and the bands for all channels,
including the total power channels, are listed in table 1. For the
correlator channels these bands include the effects of phase
decoherence.

\vspace{1cm}

\subsection{Receiver Noise and Sensitivity}

Three methods were used to measure POLAR's noise temperature.
Y-factor measurements \citep{poz90} of the total power channels
were performed with both a cooled internal calibrator (which is
accurate, but does not include the effects of the feed-horn) and
an ambient temperature external load (which is faster, but
requires a larger dynamic range) to ensure consistency. The
ambient temperature loads used were 300 K, 77 K, and the sky
($\simeq 12\,$K, zenith).

In addition to the two y-factor methods $T_{rec}$ was inferred from
noise measurements. Given the instrument bandwidth $\Delta\nu$,
the voltage fluctuations $\Delta V_{rms}$ in an integration time
$\tau$, and the calibration coefficient, $g$ in [V/K], the noise
temperature of the receiver can be estimated using the radiometer
equation: \beq T_{rec} = \frac{g^{-1}}{\kappa}\Delta
V_{RMS}\sqrt{\Delta\nu \tau} - T_{load}, \eeq where $\kappa=1$ for
the total power channels, and $\kappa = \sqrt{2}$ for the
correlator channels. The noise equivalent temperature (NET) of the
radiometer is related to the RMS temperature fluctuations via
$\Delta T_{rms}=\rm{NET}/\sqrt{\tau}$. For both the total power
channels and the correlators $\Delta T_{rms}$ is a linear function
of the load temperature. The x-intercept of these lines is equal
to the negative of the system noise temperature.

The system noise temperature is dominated by the noise temperature
of the HEMT amplifiers which are $\simeq 30$ K for both
amplifiers. However, the contribution of the room-temperature
amplifiers, as well as loss in components preceding the HEMTs,
cannot be neglected. The dominant lossy elements preceding the
HEMTs are the cryogenic isolators and the dewar's vacuum window.
The isolators' exact physical temperature is unknown, but
estimated to be $\lesssim 40$K, which is the physical temperature
of the horn, so this is a worst-case estimate. Their insertion
loss is 0.1 dB. The loss of the vacuum window is conservatively
estimated at 1\%. The room temperature RF amplifiers have noise
temperatures of $232$ K. The total estimated system noise
temperature including all factors is $T_{sys}=46$ K. Table
\ref{t:polarspex} displays the noise temperatures of all channels
using the linear intercept (noise) method outlined above. Some compression
was discerned for the highest ambient temperature load used (300 K).
Correlator J3 exhibits the highest level of compression, has the largest OMT insertion loss,
has the smallest bandwidth and obtains the highest NET.

The three methods used to estimate the receiver noise temperature
of the total power channels (internal load, external load and
noise fluctuation method) agree to within $\sim 5$ K. In the
field, the external load method was used to track the noise
temperatures on a daily basis. Using liquid nitrogen and the sky
for the loads, no compression was observed in any channel. For the
correlator channels, only the noise temperatures estimated from
the noise method were used. Noise temperatures for all
channels are displayed in table 1.

\section{Calibration}
\label{s:calib} A calibration accurate to $\simeq 10\%$ was deemed
necessary for POLAR given the expected signal levels at large
angular scales. This goal was achieved for all correlator channels
with an absolute calibration method. An ideal calibration source
would be a polarized astrophysical point source with enough power
to be seen in ``real-time''. For illustration, we compute the
power needed to produce a 5$\sigma$ detection in a 1 second
integration -- bright enough to detect in real-time. The antenna
temperature seen by \polars\ total power detectors when viewing a
source of flux density $S_\nu$ is $T_{ant} = 2.8 \, S_\nu\,
\frac{\mu\rm{K}}{\rm{Jy}}.$ \polars\ NET $ \simeq 1$ mK $s^{1/2}$
, so a source of antenna temperature $T_{ant} \simeq 5$ mK is
required for a $5\sigma$ detection in one second. This is
equivalent to a $\sim1700$ Jy source. For comparison, Cas-A, the
brightest known radio source, has a flux density of only $194\pm
5$ Jy at 32 GHz \citep{mas99}. Since Cas-A is less than 10\%
polarized at 32 GHz, the polarized signal is smaller still.
Clearly, no astrophysical sources were suitable for \polars\
calibration. In addition, the rotation mount is not pointable, so
\polar\ can only observe sources at zenith transit. Instead,
polarized signals were created by reflection of black-body
emission from wire grids (in-laboratory calibration) and
dielectric sheets (during the observing campaign).
\subsection{Wire Grid Calibrator}
Two methods of calibration were used, depending on the dynamic
range required for the measurement. Initially, a wire grid was
used to test the receiver in the lab and to probe instrumental
polarization and cross-polarization behavior. The grid produces a
highly polarized ($>99\%$), bright ($T_{\rm{ant}}=200$ K) signal.
The limited dynamic range of the receiver does not allow the wire
grid to be used as a calibrator when the instrument is in its
observing (highest gain) configuration. However, the grid was
extremely useful for characterizing the polarimetric fidelity of
the receiver.

Wire grid calibrators (WGC) are useful for near field polarization
calibration \citep{chu75,lub81,gas93}. The WGC produces correlated electromagnetic
fields in each arm of the receiver, and is placed outside the cryostat for rapid
implementation. The grid (figure \ref{f:window}) transmits
thermal radiation from a black-body source in one polarization,
and reflects thermal radiation from a second black-body source (at
a different temperature) into the orthogonal polarization. For the
\polar\ calibrator, the cold load is located above the grid and
produces electric fields $E_1$ and $H_1$, and the warm load
produces fields $E_2$ and $H_2$. $E$ and $H$ refer to the
orthogonal electric field components produced by the two loads.
Ideally, $H_1$ is transmitted and $E_2$ is reflected into the
feed-horn producing a $\simeq 100\%$ polarized diffuse source with
an antenna temperature approximately equal to the thermodynamic
temperature difference between the two loads.

The wire-grid calibrator was fabricated by deposition of copper
onto a 24 inch $\times 24 $ inch $times$ 0.002 inch mylar
substrate. The wires are 0.008 inch wide with 0.008 inch pitch.
For support the grid is sandwiched between Dow Corning ``pink"
Styrofoam sheets (emissivity $\le 1\%$), and the sandwich is
mounted at $45\arcdeg$ to the aperture plane (figure
\ref{f:window}). The grid has an integrated bearing system which
allows it to rotate directly over the vacuum window and feedhorn.

\begin{figure}[b]
\hspace{-.0cm}
\centerline{\epsfxsize=10.5cm\epsffile{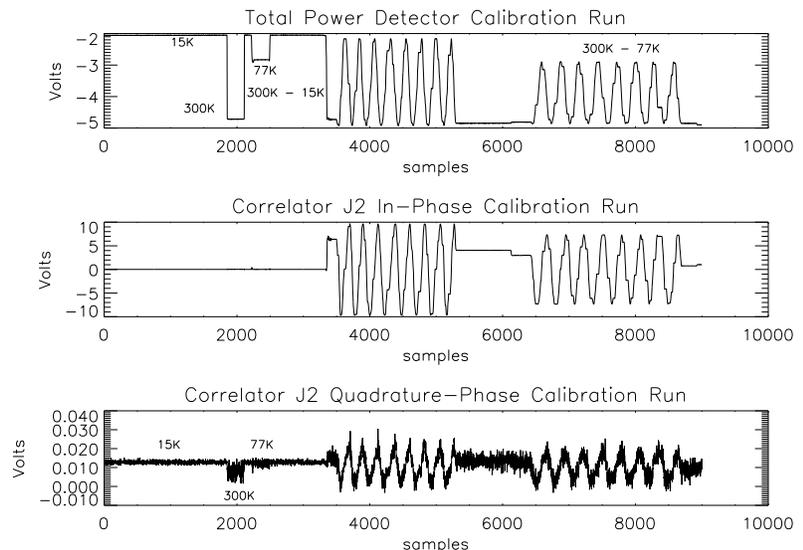}}\caption{Calibration
Run for Correlator Channel J2 and TP-E. Voltages out of correlator
J2 and total power detector TP-E (top figure) during calibration
with the wire grid calibrator are displayed. Output from TP-E is
$90\arcdeg$ out of phase with respect to the correlator channels.
TP-E uses a negative polarity total power detector. The middle
figure shows the voltage out of J2's in-phase lock-in detector,
the bottom figure show the corresponding voltage out of J2's
quadrature phase lock-in detector. The various temperature loads
are indicated at the time they are applied. The first set of
oscillations corresponds to a polarized signal obtained using a
300 K load (reflected) and the sky (transmitted), which produces a
$\sim 260$ K signal. The second set of oscillations corresponds to
a polarized temperature obtained by using a 300 K load and a
liquid nitrogen load producing a $\sim 190$ K signal. The output
of the quadrature phase detector is suppressed by $\sim 30$ dB
relative to the in-phase channel. The noise envelope of the J2 QPC
detector is a function of the load temperature since the
correlator acts like a negative polarity power detector and is
thus useful as a noise monitor. \label{f:calibrationrun}}
\end{figure}

The correlator output voltage depends on the coherence of the
electric fields produced by the thermal radiators. However, only
the antenna temperatures of the hot and cold loads are known, not
the electric fields produced in the $x$ and $y$ directions.
Fortunately, as shown below, only the antenna temperatures are
needed. The field input to the feed-horn is $\vec{H_1}+\vec{E_2}$.
In terms of the $(x',y')$ basis of the feed-horn and OMT and the
$(x,y)$ basis of the rest frame of the WGC, the magnitude of the
electric fields $E$ and $H$ produced by the WGC as the grid is
rotated (about the vertical) with respect to the polarimeter by an
angle $\alpha$ is: \beqa
& E_{x'} = E_x \cos \alpha + E_y\sin \alpha\nonumber \\
& E_{y'} = -E_x\sin \alpha + E_y\cos \alpha . \non \eeqa Since the
load fills the antenna far-field beam (edge taper $> 20$ dB) , the
output of the correlator from the coherence function given by
equation \ref{e:coherence} is: \beqa
& V_{out}\propto \lan \hat{E}_{x'}(\nu)  \tilde{E}^*_{y'}(\nu)\ran \non\\
& = \lan (E_x \cos \alpha + E_y \sin \alpha)(-E^*_x \sin \alpha +
E^*_y\cos \alpha)   \ran, \non. \eeqa Performing the
multiplication, we obtain: \beqa V_{out} \propto  \lan
-E_{x}\cos\alpha E^*_{x} \sin \alpha +
E_{y}\sin \alpha E^*_{y}\cos\alpha\ran \nonumber \non \\
=\lan E_{y} E_{y}^* - E_{x} E^*_{x} \ran \sin\alpha \cos\alpha
\non
\\=Q \sin 2\alpha =  \gamma (T_y - T_x) \sin 2\alpha \non \eeqa
where $\gamma$ converts antenna temperature (measured by the
radiometer) to intensity (the units of the Stokes parameter, $Q$).
Note that at $\alpha = 0\arcdeg, 90\arcdeg, 180\arcdeg,
270\arcdeg$ the correlators have zero output as the fields are
completely aligned along one axis of the OMT. Ideally, the grid
would reflect $T_{hot}$ from the side in 100\% horizontal
polarization and transmit $T_{cold}$ from the top in 100\%
vertical polarization, resulting in \beq V_{out} \propto \gamma
(T_{cold} - T_{hot}) \sin 2\alpha\label{e:wgccalout}.\eeq In
practice, due to loss and reflection, the following antenna
temperatures are observed at the feed-horn in the two orthogonal
polarizations \citep{gas93}: \beqa\label{e:twopoltemps} T_{hot'} =
r_{||} [ (1-r_{l})T_{hot} + r_lT_{bg}] + \hspace{2cm}\non
\\(1-r_{||})[(1-r_l)T_{cold} + r_l T_{bg}]\\
T_{cold'} = t_\perp [ (1-r_l)T_{cold} + r_l T_{bg}] + \hspace{2cm} \non\\
(1-t_\perp)[(1-r_l)T_{hot} + r_l T_{bg}],\non \eeqa where $r_{||}$
is the grid reflection coefficient for radiation polarized
parallel to the wires, $t_{\perp}$ is the grid's transmission for
radiation polarized perpendicular to the wires, $r_l$ is the
reflection coefficient of the load, and $T_{bg}=T_{hot}$ is the
effective background temperature surrounding the calibrator. In
the above equations, the effects of the emissivities and
dielectric constants of the mylar and Styrofoam have been
neglected.

Two pairs of temperature differences were used to characterize
\polar. Using a 300 K load (in reflection) and the sky (in
transmission) a polarized antenna temperature of 256 K is
obtained. Using a 300 K load (in reflection) and a 77 K Liquid
Nitrogen load (in transmission) an antenna temperature of 196 K is
obtained. The following properties of the grid were calculated
using Fresnel's equations and measured indices of refraction
(\cite{gol98} and references therein): parallel polarization
reflectivity $r_{||} = 0.995$, perpendicular polarization
transmissivity $t_\perp= 0.95$, load reflectivity $r_{l}= 0.02$. A
plot of a calibration run is shown in figure
\ref{f:calibrationrun}.

\subsubsection{Gain Matrices}
Following \citet{gas93} the output of the polarimeter versus
rotation angle is modelled as a linear combination of the Stokes
parameters at the feed horn. The three output voltages from the
two total power detectors and a correlator are modelled as a
vector: \beq \textbf{v} = \hat{\textbf{g}} \bf{T_f}+\textbf{o} +
\textbf{n} \label{e:gainmatrix} \eeq

where $\bf{\hat{g}}$ denotes the $3\times3$ gain matrix,
$\bf{T_F}=(T_c,T_h,T_c-T_h)$ is the vector of input antenna
temperatures produced by the grid, and $\bf{o}$ and $\bf{n}$
represent offset and noise contributions to $\bf{v}$ respectively.
The coordinate basis of $\bf{v}$ is defined by the OMT. For
simplicity we only consider the coupling of one correlator to the
total power channels, so the dimension of the system is 3 rather
than 5 for POLAR. Ideally, $\hat{\bf{g}}$ would have only on-diagonal
elements, however the off-diagonal elements of $\hat{\bf{g}}$
correspond to various non-idealities of the instrument which will
result in offsets in our measurements.

As the grid rotates, the resulting vector of voltages is
recorded and a least-squares fit is made to the data using the
radiometer model of equation \ref{e:gainmatrix}. The gain matrix
parameters, including the off-diagonal cross-talk elements, and
the offsets are recovered for each calibration run. With the
antenna temperatures of the loads given by equation
\ref{e:twopoltemps}, the voltages out of the two total-power
channels and the correlator channel are:
\beqa\label{e:gainmatrix2} \bf{v} =\pmatrix{ g_{yy}T_c +
g_{yx}T_{s} + g_{yQ}(T_{hot'} - T_{cold'})\sin 2\alpha + o_y\cr
g_{xy}T_c + g_{xx}T_{s} + g_{xQ}(T_{hot'} - T_{cold'})\sin 2\alpha
+ o_x \cr g_{yQ}T_c + g_{xQ}T_{s} + g_{QQ}(T_{hot'} -
T_{cold'})\sin 2\alpha + o_Q  } + \bf{n} \non
\\\hspace{10pt} \eeqa where $$T_c = T_{hot'} \cos ^2 \alpha +
T_{cold'} \sin ^2 \alpha$$ and $$T_{s} = T_{hot'} \sin ^2 \alpha +
T_{cold'} \cos^2 \alpha.$$

To recover $\bf{\hat{g}}$, we first integrate long enough that the
noise term, $\bf{n}$, is negligible, and then average the offsets,
$\bf{o}$, as a function $\alpha$ and subtract them. Then equation
\ref{e:gainmatrix2} is inverted to obtain $\bf{\hat{g}}$. The
on-diagonal elements ($g_{xx},g_{yy},g_{QQ}$) of $\bf{\hat{g}}$,
dominate the matrix; they are the terms which measure the system
calibration in [V/K]. Typical values are $\sim (100 \;
\rm{K/V})^{-1}$. The off-diagonal elements encode the system's gain imbalance, cross-talk, and
imperfect isolation between polarization states. The $g_{xy}$
terms are approximately $1\%$ of the $g_{xx},g_{yy}$ terms, and
the $g_{xQ},g_{yQ}$ terms are $\le 1\%$ of the $g_{QQ}$ terms for
all three correlators.

There are two classes of systematic effects which lead to the
off-diagonal elements $g_{xy}=g_{yx}$ and $g_{xQ}=g_{yQ}$. To
analyze the effects of $g_{xy}\neq 0$,  we set $g_{yQ}=n_y=o_y=0$,
and identify the first non-ideality, $g_{xy}$ (which is equal to
$g_{yx}$). This implies that at $\alpha = 0$, when only $T_{hot'}$
should be observed, $v_y = g_{yy}T_{hot'}+g_{yx}T_{cold'}$ is
observed. Thus, $g_{xy}$ terms represent cross-polarization. The
main contribution to the correlator offset is from
cross-polarization of the OMT and/or imperfect isolation of the
OMT. The off-diagonal elements, \eg, $g_{xQ}$, are attributed to
gain differences in the feed horn's E and H plane power response,
and can be equalized in hardware or software.

Since two pairs of temperature differences ($300$ K load vs. $77$
K load and $300$ K load vs. the sky) were used, the calibration
constants as a function of the temperature difference were
measured and checked for linearity. The two pairs of loads
produced effective polarized antenna temperatures of 256 K and 196
K, and it was verified that the calibration constants were equal
to better than 10\% over this range for J1 and J2.

\subsection{Dielectric Sheet Calibrator}
As mentioned above, calibrations performed during the observing
campaign did not use the wire grid calibrator. The primary reason
for this was the limited dynamic range of the polarimeter; both
the last stage of IF amplifiers and the correlators themselves
began compressing when the antenna temperature was $\sim 100$ K in the observing (high gain) configuration.
When the sky was the cold load, a full rotation of the wire grid
produces a modulated signal with amplitude
$100\,\rm{K}<T_{ant}<250\,\rm{K}$. The variation in bias power to
the correlators produced by the WGC as it was rotated was
significant. The largest imbalance loaded the correlators with 40
K on one port and up to 290 K on the other port. This imbalance is
undesirable and was the initial reason the Dielectric Sheet Calibrator
(DSC) was used \citep{ode01}. During observations, the calibrator
should produce a total power load similar to the sky loading,
which is only slightly polarized.

During the observing campaign, the wire grid calibrator was
replaced by a thin (0.003 inch) polypropylene film. This produces
a signal that is only partially polarized. The polarized signal
produced by the (DSC) is \beqa\label{eq:DSC} Q = [(T_{hot} -
T_{cold})(R_{TE} - R_{TM}) + \\(T_S - T_{cold})(\epsilon_{TE} -
\epsilon_{TM}) ] \cdot \sin{2\alpha}\non \eeqa where $T_S$ is the
physical temperature of the dielectric sheet. $\epsilon_{TE}$ is
the emissivity of the dielectric in the TE polarization state
(perpendicular to the plane of incidence), and $\epsilon_{TM}$ is
the emissivity in the TM polarization state (parallel to the plane
of incidence). Note that this expression reduces to equation
\ref{e:wgccalout} in the wire grid case, where $R_{TE} - R_{TM} =
1$, and $\epsilon_{TE}=\epsilon_{TM}=0$.

The reflection coefficients of the DSC is determined by the
dielectric constant and the geometry:
\begin{equation}\label{Rgeneral}
R = \frac{\left[\cos^2{\theta} - \gamma_i^2\right]^2 \
\sin^2{\delta}}
        {4 \gamma_i^2 \cos^2{\theta} \cos^2{\delta} \ + \
\left[\cos^2{\theta}+\gamma_i^2\right]^2 \sin^2{\delta} }
\end{equation}
where $i\in\{TE,TM\}$, and
\begin{eqnarray}\label{gammaTE}
\gamma_{TE} & \equiv & \sqrt{\ n^2 - \sin^2{\theta}} \\
\label{gammaTM}
\gamma_{TM} & \equiv & \frac{1}{n^2} \sqrt{n^2 -\sin^2{\theta}} \\
\delta & = & 2\pi \nu t \sqrt{n^2 - \sin^2{\theta}}\,,
\end{eqnarray}
where $n$ is the refractive index of the sheet, $\nu$ is the
frequency, $t$ is the sheet thickness, and $\theta$ is the angle
of incidence of the incoming radiation.  For our geometry, $\theta
\ = \ 45^\circ$.

For 0.003 inch polypropylene at 30 GHz, $R_{TE} - R_{TM} \approx
0.2\%$. The emission from the sheet is $\sim 4$ mK per 0.001 inch
of thickness, and is negligible compared to the reflection-induced
signal. When the sky is used as the cold load and a 300 K hot load
is used, $T_{hot} - T_{cold} \approx 260 K$, and produces a
rotation modulated polarized calibration signal of $\sim 500$ mK,
and an unpolarized background power of $\sim 10$ K (the sky
temperature).

\begin{figure}[t]
\hspace{-.0cm}
\centerline{\epsfxsize=8.5cm\epsffile{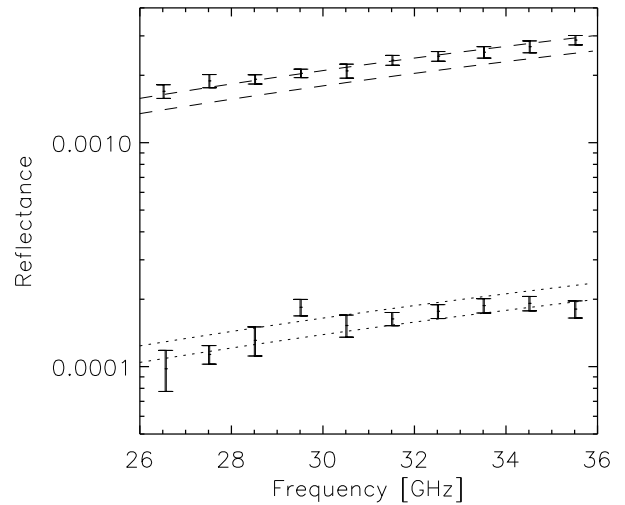}}\caption{Comparison
between laboratory reflectivity measurements and theory on 0.003
inch polypropylene situated at $\theta = 45^\circ$. Errors in the
data are mostly systematic, arising from standing waves in the
system. The uncertainty in the model is due to both thickness
variations and uncertainties in the index of refraction of the
dielectric sheet. $R_{TE}$ corresponds to the upper set of curves
(dashed), and $R_{TM}$ to the lower set of curves
(dotted).\label{f:dielectric_sheet}}
\end{figure}

Equation \ref{Rgeneral} was verified in laboratory tests; the
results for 0.003 inch polypropylene are given in Figure
\ref{f:dielectric_sheet}. The primary sources of error in our
final calibration were uncertainties in the indices of refraction
and the slight thickness variations in the sheet; these 5\%
variations lead to final calibration error of $8.5\%$ for J1 and
J2, and $11\%$ for J3.

\section{Systematic Effects and Radiometric Offsets}
\label{s:systematics}
\subsection{System Sensitivity Degradation}

Once the conversion between voltage and temperature is known, by
measuring the voltage RMS the temperature RMS can be obtained. The
noise in an arbitrary integration time, $\tau$, is $\Delta T_{RMS}
= \rm{NET}/\sqrt{\tau}$. The most naive technique to obtain the NETs is simply
to calculate the RMS of the time stream in a one-second segment
and convert from voltage to temperature. This approach, however,
over-estimates the NET, and only applies when the noise is white (no $1/f$ noise.
A general expression for the post-detection spectral
density of correlation and total power radiometers which includes
the effects of gain fluctuations, $\Delta G(f)$, a system offset,
$T_{offset}$, and offset fluctuations, $\Delta T_{offset}(f)$, is
given by \citep{wol95,car01}:

\beq \label{eq:corrtruenet} P^{\rm{corr}}(f) =
2\frac{\kappa^2T_{sys}^2}{\Delta \nu}+T_{offset}^2\Delta
G^2(f)+\Delta T^2_{offset}(f), \eeq

\beq \label{eq:tptruenet} P^{\rm{TP}}(f) =
2\frac{T_{sys}^2}{\Delta \nu}+T_{sys}^2\Delta G^2(f)+\Delta
T^2_{sys}(f). \eeq

\begin{figure}[t]
\hspace{-.4cm} \centerline{\epsfxsize=10.cm\epsffile{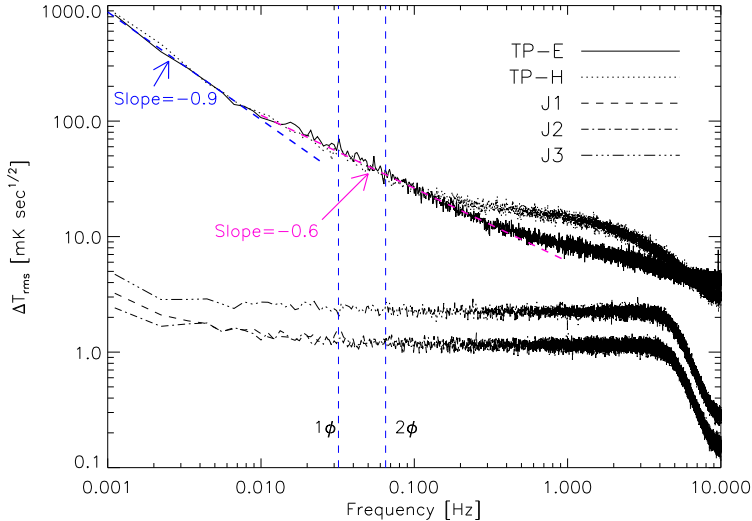}}
\caption[Power Spectra of All Correlator Channels and total power
detectors] {Square-root power spectra of all five \polar\ signal
channels. The $1/f$ behavior of the total power detectors, and the
low-pass anti-aliasing filters are evident. The CTI coldhead
expansion/compression cycle is at 1.2 Hz, and no contamination is
observed in the signal channels. Vertical lines indicate the
rotation frequency ($1\phi$) and the Stokes parameter modulation
frequency ($2\phi$). The low frequency rise in the total power
detector spectra is due to both HEMT gain fluctuations (for $f >
0.01$ Hz) and atmospheric fluctuations (for $f < 0.01$ Hz). A
``$1/f$" fluctuation spectrum has a slope of -0.5 on this plot and
a Kolmogorov atmospheric fluctuation spectrum has a power law
slope of $\simeq -1.3$. The low frequency rise in the correlator
spectra at $f<0.01$ Hz has a power-law slope of $\sim -1$ which
indicates that it arises from atmospheric fluctuations. Correlator
J3 has a smaller bandwidth and higher isolator loss than J1 or J2
leading to a higher white noise level. The anti-aliasing filter on
TP-H had a low-pass cutoff at 5 Hz (identical to the correlators),
while TP-E's low-pass cutoff was at 50 Hz leading to different
spectral shapes. \label{f:allpsd}}
\end{figure}

Note that the second and third terms of equations
\ref{eq:corrtruenet} and \ref{eq:tptruenet} do not depend on the
RF bandwidth, $\Delta \nu$ and do not, in general, integrate down
with time. The audio frequency, $f$, dependence of the gain
fluctuations for the HEMT amplifiers is $\Delta G(f)\propto
f^{-1}$ \citep{wol95}. These equations, along with Figure
\ref{f:allpsd} (which shows the power spectra of all three
in-phase correlator channels and both total power detectors during
an observation run) illustrate the relative performance tradeoffs
of the total power polarimeter versus the correlation polarimeter.
\polar\ uses both types of radiometer; however, the total power
polarimeter channel is used only as an atmospheric monitor. The
instantaneous difference between the two total power channels
(TP-E and TP-H) is proportional to the Stokes Q parameter in the
OMT frame, and after $45\arcdeg$ rotation would provide the U
parameter. However, for a total power receiver the HEMT gain
fluctuation noise $\Delta G(f)$ in equation \ref{eq:tptruenet}
multiplies the \emph{system} temperature
($T_{sys}=T_{rec}+T_{ant}$) rather than the \emph{offset}
temperature as in equation \ref{eq:corrtruenet}. This produces the
dramatic $1/f$ rise in the total power detectors' PSD, which is
greatly diminished for the correlator channels. This allows us to
slowly modulate the signal by rotation of the radiometer at 33
mHz, rather than at several Hz as would be required for the total
power channels. It is clear from the spectra that the correlators
are far more sensitive and stable than the total power detectors.

\begin{figure}[b]
\hspace{-.4cm}
\centerline{\epsfxsize=10.cm\epsffile{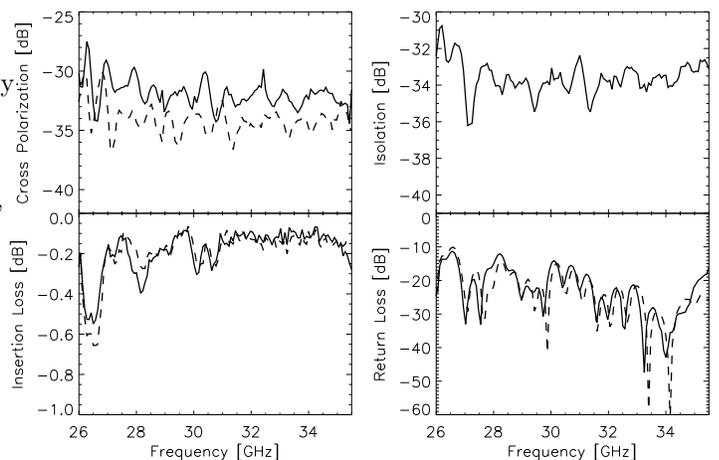}}\caption{Properties
of \polars\ OMT. Isolation, cross-polarization, return loss, and
insertion loss are shown across the $\ka$ band. For the insertion
and return loss plots, the solid line is for the E-plane of the
OMT and the dashed is for the H-plane. For the cross polarization,
the solid line is is E-plane input, H-plane output and the dashed
line is H-plane input, E-plane output. All properties improve at
the highest RF frequencies, leading to decreased spurious
polarization for channels J1 and J2 relative to J3.
\label{f:omtquad}}
\end{figure}

The correlation radiometer offset is produced by signal power
which is correlated between the two polarization states. This
effect is primarily the result of non-zero cross polarization and
polarization isolation of the OMT. The total spurious polarization
generated by the OMT is due to both cross polarization and
imperfect isolation. In Figure \ref{f:omtquad} the isolation,
cross polarization, insertion loss, and return loss are shown.

Since the correlation polarimeter offset is produced mainly by
spurious polarization of the OMT, the dominant source of offset
fluctuations will be from fluctuations in the antenna temperature
of observed sources, primarily (unpolarized) atmospheric emission:
$T_{offset}(f) = {\rm SP}_{omt} T_{atm}(f)$, where $\rm{SP}_{omt}$
is the OMT's spurious polarization (sum of the isolation and
cross-polarization). The OMT's cross-polarization dominates the
spurious polarization, since the total isolation between the
polarization states is the sum of the isolation of the OMT and
cryogenic isolators on the output ports of the OMT. The total
isolation is $< 50$ dB. In addition, path length differences in
the two arms leads to phase decoherence for signals reflecting off
the HEMT amplifiers and propagating in the reverse direction. This
reduces the effects of non-zero isolation to negligible levels.
The atmospheric component of antenna temperature fluctuations at
30 GHz follows a Kolmogorov spectrum which falls as $T_{atm}(f)
\propto f^{-8/3}$ \citep{car01}. If the experiment is not
modulated at a frequency much higher than the knee frequency of
the fluctuation spectrum, these terms will dominate the system
NET.  To perform this modulation \polar\ was rotated at $0.033$
Hz. Figure \ref{f:allpsd} shows the power spectra produced by all
radiometer channels, including the effects of atmospheric
fluctuations.

The stability of the offsets over a single rotation of the
instrument is crucial to the recovery of the Stokes parameters.
Note that the behavior of the noise should be independent of the
phase of the reference waveform supplied to the phase-modulation
lock-in detectors. The QPC are insensitive to correlated signals,
including uncorrelated atmospheric emission which is spuriously
correlated by the OMT. The QPC therefore show almost no $1/f$
noise; residual $1/f$ noise at the $\simeq -25$ dB level is due to
cross-talk in the low frequency electronics and the inability to
perfectly match the phase in each arm.  The QPC proved to be
powerful monitors of the intrinsic noise of the radiometer. Over
the course of the 2000 observing campaign, we found periods of
high offset and large offset fluctuations to be correlated with
environmental effects, especially the occasional formation of dew
and ice on the vacuum window.

\subsection{Optical Cross Polarization}
The corrugated scalar feed horn demonstrates low
cross-polarization\citep{cla84}. However, even for an ideal and
completely symmetric feed there is always non-zero
cross-polarization. For an ideal horn, the cross-polarization
induced by scattering in a plane containing the polarization axis
is identically zero since there has been no polarization
conversion. This is also manifestly true for scattering in a plane
perpendicular to the polarization axis. However, using a simple
geometric optics approximation it can be demonstrated
\citep{cla84} that there is polarization conversion
(cross-polarization) which varies as $\sin^2\phi$, and will be
peaked at $\phi = 45\arcdeg, 135\arcdeg, 225\arcdeg, 315\arcdeg$
where $\phi$ is the azimuthal angle in the aperture plane. The
maximum cross-polarization of the feed was measured to be $\le
-40$ dB (see figure \ref{f:crosspolbeammap}). It is primarily the
off-axis response in the near side-lobes that show
cross-polarization. As shown in \citet{car01}, the quadrupolar
anisotropy on scales comparable to the FWHM is the dominant source
of spuriously correlated response by the feed. \polars\ vertical
drift scan geometry and low side-lobe level reduced the effect of
cross-polarized optical response to negligible levels.

\subsection{Non-Ideal Correlation Radiometer Behavior}
\label{ss:phaseshift} The most significant non-ideal behavior of the
correlation radiometer results from electrical path length
mismatch between the input arms. From \eq{e:coherence} the
correlator's DC output is proportional to $\cos (\Delta\phi_\nu)$
where $\Delta\phi_\nu$ is the phase shift between the two arms of
the radiometer. A $90\arcdeg$ phase shift therefore results in a
zero signal-to-noise ratio. The path length difference $\Delta L$
introduces a dispersive phase shift, $\Delta\phi(\nu,\Delta L)$.
Recalling that POLAR measures the cross-correlation at $\tau=0$
lag, and assuming constant power spectra across the RF band for
the source, beam, and radiometer transfer functions, from
\eq{e:coherence} we have:

\beq R(0) \propto
\int_{-\pi}^{+\pi}\tilde{\gamma}(\theta)\tilde{B}(\theta)\vert\tilde{H}\vert^2\,d\theta
\int_{\nu_o}^{\nu_o + \Delta\nu_{RF}} \cos \Delta\phi \; d\nu .
\eeq The contribution of each spectral component is thus weighted
by the cosine of its phase. It is therefore
imperative to accurately match the path lengths in the system. To
determine the phase mis-match a completely polarized signal is
injected into the OMT input. The injected signal is swept in
frequency across the RF band. By measuring the frequency
modulation of the correlator spectrum by the $\cos\Delta\phi$
envelope, the equivalent path length imbalance can be
determined. The path difference measurements agreed
with measurements of the physical waveguide path difference. To
balance the path lengths, sections of waveguide were added to the
shorter arm of the receiver. Comparing the theoretical NET given
the RF bandwidth to the measured NET, we estimate that the
differential phase shift for the correlation channels was $<
20\arcdeg$.

The remaining non-ideality results from gain asymmetry between
arms, across the band passes. These effects can be caused by
mismatched bands, temperature dependence, and phase instability of
the amplifiers and/or the correlator. In practice it is impossible
to eliminate all such effects. Following \citet{tho98}, in table
\ref{t:freq} estimates of the tolerable level of a few effects
which could contribute to a 2.5\% degradation of the
signal-to-noise ratio of the correlation receiver.

\subsection{Polarimetric Offsets}
\begin{figure}[t]
\hspace{-.2cm}
\centerline{\epsfxsize=10.2cm\epsffile{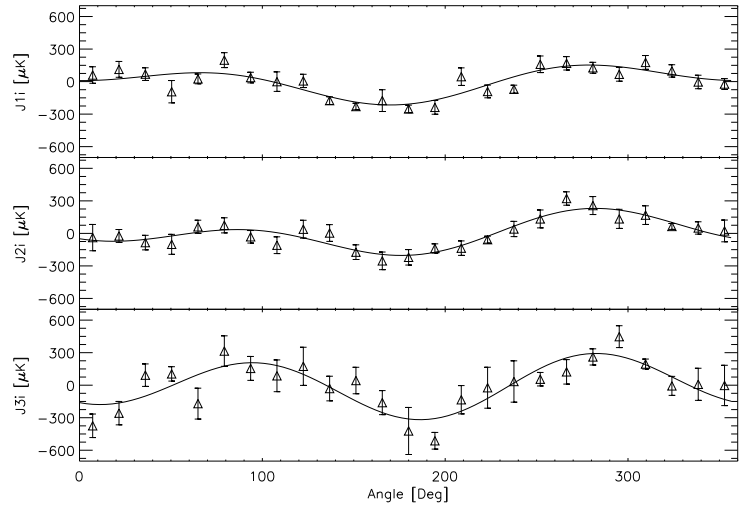}}\caption{Correlator
outputs binned into rotation angle for 206 rotations (1 hour 43
minutes) of data obtained during observations on 2 May 2000. The
DC offset ($I_o \simeq 100$ mK) of each channel has been removed.
Unpolarized offsets and Stokes parameter offsets are visible for
each of the three correlator channels. The relative phasing and
offset magnitudes are consistent with the model presented in
sub-section \ref{s:offsets}. Fits to the Stokes parameter offsets
are presented in table \ref{t:offsets}. \label{f:offsets}}
\end{figure}
\label{s:offsets} In the analysis, for each rotation of the
polarimeter, the correlator outputs are binned into rotation angle
$\theta_t$ and fit to
\beq I(\theta_t) = I_o + C\cos\theta_t + S\sin
\theta_t + Q\cos 2\theta_t + U\sin 2\theta_t.\label{e:signal}\eeq where $\theta_t = 2\pi f t$ and $f=0.033$ Hz.
In addition to the Stokes parameters $Q$ and $U$, the terms $C$ and $S$ are monitored to
determine our sensitivity to rotation-synchronous systematic
effects, and to monitor atmospheric fluctuations. Phase sensitive
detection at twice the rotation frequency removes $I_o$
and other instrumental effects that are not modulated at this
frequency.

Figure \ref{f:offsets} shows the output of 206 co-added rotations (1 hour
43 minutes of data) from the night of 2 May 2000 binned into a
single rotation to increase the signal to noise ratio of the
systematic effect. Fits to $I_o\,,C\,,S\,,Q\,,U$ for these plots
result in the Stokes parameter offsets for each channel in the
instrumental coordinate system.

The cause of the offset was
carefully investigated. Initially, the magnetic coupling of the
cryogenic isolators to the earth's field was suspected. Helmholtz
coils were used to produce a field of $\simeq 10$ Gauss at the
position of the isolators. The offsets remained unchanged after
one hour of integration with the coils in place. The coils were
then located at four other azimuthal positions around the dewar
and no observable effects were noticed.

The modulated signals were found to be consistent with a common
\emph{optical} offset for each channel (as indicated by the
consistent phasing of the signals across the channels).
Unpolarized flux is correlated in the receiver due to the OMT's
cross polarization and imperfect isolation, causing $I_o \neq 0$.
The optical flux is believed to be unpolarized, but anisotropic, with a
dipolar and quadrupolar dependence on the rotation angle
$\theta_t$, producing spuriously polarized components $Q$ and $U$.
The quadrupole anisotropy is most likely caused by the outer,
reflecting, ground screen, which is a square `scoop' centered on
the dewar axis, while the dipole anisotropy is attributed to the
feed horn being located $\simeq 3$ inches radially outward from
the dewar axis centerline. The dipolar ($S$ and $C$) and
quadrupolar ($Q$ and $U$) components are present at levels that
are 30 dB lower than the unpolarized offset $I_o$. The frequency
dependence of the offset is consistent with the OMT's performance.
Both the cross-polarization and the isolation of the OMT degrade
with decreasing RF frequency. Therefore, since the radiometer
offset is primarily due to the OMT's cross-polarization, the
offset will be largest for J3 since it is the lowest frequency
band. Atmospheric emission that is truly polarized by the
groundscreen would have a spectrum that increases with frequency,
contrary to what was observed.

Table \ref{t:offsets} presents the
offsets as a function of channel for the 2 May 2000 data. The offset phase angle
dependence was roughly constant throughout
the season, whereas the magnitudes of the offsets were correlated
among channels and varied with observing conditions; most notably
humidity and atmospheric opacity. This again supports the
hypothesis that unpolarized anisotropic atmospheric flux is polarized
by the OMT.

To be used in the cosmological data analysis, the magnitude and
phase of the offset must be stable over $> 4$ hour timescales.
Approximately 50\% of the surviving sections of data have stable
offsets for $>8$ hour periods of time. Our sensitivity to Stokes
parameter offsets is minimized by constraining the demodulated
data to have no dependence on an overall Stokes parameter offset.
This is a generalization of the procedure outlined in
\citet{bjk98} to treat (unpolarized) offsets for CMB temperature
anisotropy experiments. The offset removal procedure not only
constrains the final maps produced to have zero offset, but also
accounts for the sensitivity degradation induced by the offset
removal. The offsets are computed from maps produced for each
channel for each contiguous 3 hour block of data that survives the
data editing criteria (denoted as a ``section''). The offset for
each section is determined by enforcing consistency (within the
error bars) between the maps constructed from all sections
measuring the same pixel on the sky. This induces correlations
between sections of data, and between adjacent pixels mapped in
the same section. The offset removal and data analysis procedure
are discussed in detail in ~\citet{ode02}.

\section{Meteorological Effects and Data Selection}
\label{s:cuts} A variety of weather-related phenomena was
encountered during the Spring 2000 observing campaign. We compiled
data on the Pine Bluff, Wisconsin area from both the National
Weather Service and the GOES-8 satellite data served by the Space
Sciences and Engineering Center at the University of
Wisconsin-Madison\footnote{\texttt{http://www.ssec.wisc.edu}}.

\subsection{General Atmospheric Effects}
\begin{figure}[t]
\hspace{-.0cm}
\centerline{\epsfxsize=9.8cm\epsffile{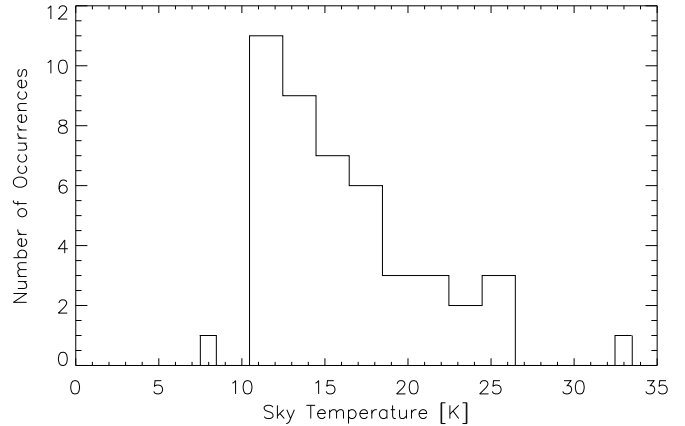}}\caption{Daily
atmospheric zenith antenna temperature distribution for the Spring
2000 observing campaign. \label{f:tatm}}
\end{figure}

The GOES-8 data are recorded hourly, and measures a $\simeq$ 5 km
by 5 km area, within 20 km of the \polar\ observatory. It provided
the cloud cover fraction of the area, and precipitable water vapor
(PWV) column height (as well as a host of other weather
variables). Periods of high PWV correlated with formation of dew
and ice on the vacuum window. Astrophysical data acquired during
these periods were not used in the analysis due to the spurious
correlation produced by reflection from the dew/ice on the window.
Cloud cover fraction exhibits a bimodal histogram, with more than
35\% of the time clasified as `totally clear' and about 15\% of
the time categorized as completely overcast. Partially cloudy days
account for the other 50\% of the distribution. \polars\ two total
power channels monitor the atmospheric zenith temperature.
Figure \ref{f:tatm} presents a histogram of daily
atmospheric zenith temperature measured over the observing season
by  \polar.

\subsection{Solar and Lunar Effects} Based on
the geometry of the inner, co-rotating, conical groundscreen, some
solar radiation will enter this screen when sunlight propagates
over the outer (fixed) groundscreen. This happens at a solar
elevation of $\simeq 10\arcdeg$. However, for this light to enter
the horn, it must scatter many times off the inner ground screen
and will then be absorbed by the inner screen's Eccosorb coating.
Below this elevation of $10\arcdeg$, solar radiation must undergo
a double diffraction to enter the system. The amount of sunlight
in the beam-pattern steeply increases as the sun rises until the
sun's elevation reaches $\simeq 49\arcdeg$, at which time
radiation from the sun can directly enter the horn. We found that,
in practice, solar contamination was undetectable below a solar
elevation of $30\arcdeg$. To be conservative, we eliminated all
data taken with the sun more than $20\arcdeg$ above the horizon.
Since data were collected 24 hours per day, this represents a
sizeable 38.6\% of our data, or $\simeq 288$ hours.

The moon is a bright microwave source, corresponding roughly to a
thermodynamic temperature of 220 K.  Its emission is dependent
upon frequency, lunar phase, and polarization.  Using the lunar
emission model presented in \citet{kei83}, the \emph{COBE} team
calculate the lunar emission in both polarization states at 30,
50, 90 GHz, and show that the polarized lunar antenna temperature
at 31 GHz is $\lesssim 1 K$ \citep{ben92}. Using a variation of
this model, and the \polar\ beam patterns, we have estimated the
polarized antenna temperature of the moon as a function of
elevation angle. During the Spring 2000 observing season the
highest lunar elevation was $68.4\arcdeg$. We removed all data
when the moon was more than $50\arcdeg$ in elevation; this
corresponds to about 1.2\% of the data, and reduces the maximum
lunar contribution to be $\ll 1\,\mu$K.
\subsection{Atmospheric Data Cut}
\begin{figure}[t]
\hspace{-.5cm}
\centerline{\epsfxsize=9.8cm\epsffile{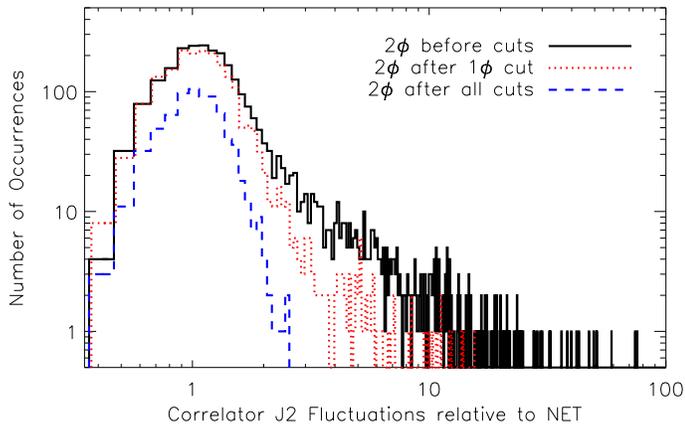}}\caption{Distribution
of correlator fluctuations for 7.5 minute averages, as a function
of cut level.  The solid (black) curve shows the fluctuation level
at twice the rotation frequency, relative to the NET of the
system, for the in-phase channel J2I, after cuts removing sun,
moon, and dew contamination have been applied.  The dotted curve
shows the same information, but after the $1\phi$ cut has been
applied. This cut removes a large number of high-level outliers,
but much of the remaining data still are contaminated. The dashed
curve shows the distribution after all cuts have been applied;
very little if any $2\phi$ contamination remains.
\label{f:twophihist}}
\end{figure}

The primary data quality cut for selecting astrophysical data is
based on the statistics of the $S$ and $C$ terms of fits to
equation \ref{e:signal}. This cut is referred to as the $1\phi$
cut. As previously mentioned, the $S$ and $C$ components are
statistically independent from the $Q$ and $U$ components. Since
the $1\phi$ component probes the power spectrum of the radiometer
at lower frequencies, it is more susceptible to contamination by
atmospheric fluctuations, and can therefore be used as an unbiased
probe of data quality which is independent of the astrophysical
data. For each 7.5 minute segment of data, the fluctuations in $S$
and $C$ are compared to 1) the expected fluctuation level from
gaussian white noise and 2) fluctuations in $S$ and $C$ from the
QPC which, as mentioned, display pure white noise power spectral
densities with amplitude equal to the radiometer NET. Figure
\ref{f:twophihist} shows the distribution of fluctuations in the
$2\phi$ component for the 7.5 minute averages for channel J2, for
the Spring 2000 observing campaign. Also indicated are the cut
levels used in the analysis presented in \citet{kea01} and the
in-phase channel (IPC) distribution after the $1\phi$ cut has been
applied.

\section{Summary}
We have described the design and performance of a novel instrument
which was recently used to set upper limits on the polarization of
the CMB. The simplicity of the optical design of the polarimeter
and the observing strategy resulted in minimal systematic effects.
Observations were conducted from a convenient location near the
University of Wisconsin -- Madison. The site was useful for work
at 30 GHz and its proximity afforded us the ability to diagnose
problems and make rapid adjustments to optimize instrumental
performance while still in the field. While no evidence for CMB
polarization was detected with \polar, the upper limits are
impressive given the brief observing season available during
Spring 2000. This is attributable to the low noise of the HEMT
amplifier front-end and the modest modulation requirements
permitted by the stable correlation radiometer back-end. \polar\ has
proven to be extremely versatile. Starting in January
2001 the \polar\ radiometer has been used as the receiver in a
search for CMB polarization at small angular scales: COsmic
Microwave Polarization at Small Scales ({\scshape{Compass}}).
Results from {\scshape{Compass}} are forthcoming and will further
demonstrate the viability of the correlation polarimeter
technique.

\acknowledgments We are grateful to a number of people who
influenced the design and analysis of \polar. Initial theoretical
guidance and encouragement came from Alex Polnarev and Robert
Brandenberger.  Dick Bond, Robert Crittenden, Angelica de
Oliveira-Costa, Wayne Hu, Lloyd Knox, Arthur Kosowsky,  Kin-Wang
Ng, and Matias Zaldarriaga provided crucial insight during the
development and analysis phases, and vastly enhanced the
scientific impact of the project. We are also indebted to several
experimentalists who worked-on, supported, or guided, the
construction of the instrument: Brendan Crill, Khurram Farooqui,
Kip Hyatt, Slade Klawikowski, Alan Levy, Phil Lubin, Melvin Phua,
Dan Swetz, David Wilkinson, Grant Wilson, Ed Wollack. BGK and CWO
were supported by NASA GSRP Fellowships. \polars\ HEMT amplifiers
were provided by John Carlstrom. This work has been supported by
NSF grants AST 93-18727, AST 98-02851, and AST 00-71213, and NASA
grant NAG5-9194.

\clearpage
\begin{deluxetable}{cc}
\tablewidth{0pt} \tablecaption{Tolerances on Correlation
Polarimeter Frequency Response Variations for a 2.5\% Reduction in
Signal to Noise Ratio (relative between arms).\label{t:freq}}
\tablehead{\colhead{Type of Variation} & \colhead{Permissable
Level}\\}\startdata
Gain Slope & 3.5 dB across band \\
Gain Sinusoidal Ripple & 2.9 dB peak-peak \\
Frequency Band Centroid Offset & 5\% of $\Delta\nu_{RF}$\\
Phase Shift Between Bands & $12.8\arcdeg$ \\
\enddata
\end{deluxetable}

\begin{deluxetable}{cccccccc}
\tablewidth{0pt} \tablecaption{\polar\ Offsets 2 May 2000. 206
Rotations (1 hour 43 minutes of data) co-added and binned
 into rotation angle\tablenotemark{\dag}. \label{t:offsets}}
 \tablehead{
& \colhead{$I_o$\tablenotemark{b}}&
\colhead{$C$\tablenotemark{c}}& \colhead{$S$ \tablenotemark{d}}&
\colhead{$Q$\tablenotemark{e}}& \colhead{$U$\tablenotemark{f}}&
\colhead{$P$\tablenotemark{g}}& \colhead{$\phi$\tablenotemark{h}}\\
\colhead{ {Channel}\tablenotemark{a}} & \colhead{ {[mK]}} &
\colhead{ {[$\mu$K]}} & \colhead{ {[$\mu$K]}}&\colhead{
{[$\mu$K]}}&\colhead{ {[$\mu$K]}}& \colhead{ {[$\mu$K]}}&\colhead{
{[degrees]}}\\} \startdata
 {J3I} &  {$133.8 \pm 1.0$}&   {$75.4 \pm 38.0$} &  {$-31.7 \pm 38.0$} &   {$-236.3\pm 38.0$} &   {$-73.6\pm 38.0$}& {$247.5\pm 47.6$} &   {$8.7 \pm 4.7$}\\
 {J2I} &  {$83.6 \pm 1.0$}&   {$76.2 \pm 20.0$} &  {$-91.3 \pm 20.0$} &   {$-125.8 \pm 20.0$} &  {$-25.9 \pm 20.0$}& {$128.4\pm 23.6$} &   {$5.8 \pm 4.6$} \\
 {J1I} &  {$88.1 \pm 1.0$}&  {$108.3 \pm 20.0$} &   {$-48.4 \pm 20.0$} &   {$-99.3 \pm 20.0$} &  {$21.0\pm 20.0$}& {$101.5\pm 23.7$} &   {$-6.0 \pm 5.8$}  \\
 {J3Q} &  {$15.76 \pm 0.09$} &   {$13.8\pm 38.0$} &  {$15.0\pm 38.0$} &   {$8.4\pm 38.0$} &  {$68.9\pm 38.0$}& {$69.5\pm 42.1$} &   {$41.7 \pm 52.5$} \\
 {J2Q} &  {$5.57 \pm 0.05$} &  {$-9.6\pm 20.0$} &   {$-25.3\pm 20.0$} &   {$29.6\pm 20.0$} &  {$-28.0\pm 20.0$}& {$41.0\pm 28.3$} &   {$-22.5 \pm 22.9$} \\
 {J1Q} &  {$7.15 \pm 0.04$} &  {$7.3\pm 20.0$} &   {$-19.1\pm 20.0$} &   {$-20.3\pm 20.0$} &  {$12.7\pm 20.0$}& {$23.8\pm 27.7$} &   {$-16.5 \pm 31.5$}   \\
\enddata
\tablenotetext{\dag}{ {refer to Fig. \ref{f:offsets} for data.}}
\tablenotetext{a}{ {`I' refers to in-phase channels; `Q' refers to
quad-phase channels.}} \tablenotetext{b}{ {Unpolarized,
unmodulated, intensity.}} \tablenotetext{c}{ {Dipole modulated
cosine term.}} \tablenotetext{d}{ {Dipole modulated sine term.}}
\tablenotetext{e}{ {Quadrupole modulated $Q$ term.}}
\tablenotetext{f}{ {Quadrupole modulated $U$ term.}}
\tablenotetext{g}{ {Magnitude of polarized offset
$P=\sqrt{Q^2+U^2}$. Over the course of the observing season, the
offsets were always\\
 in the order $P(J3)>P(J2)>P(J1)$.}}
\tablenotetext{h}{ {Phase angle of polarized offset $\phi =
\frac{1}{2}\tan^{-1}\frac{U}{Q}$. Throughout the season, the phase
angles of the offsets \\were roughly constant for the in-phase
channels. }}
\end{deluxetable}

\end{document}